\documentclass[12pt]{article}
\pdfoutput=1

\usepackage{graphics,amssymb,float}
\usepackage{graphicx}
\usepackage{color}
\usepackage{xspace}
\usepackage{placeins}
\usepackage{rotating}
\usepackage{dcolumn}
\usepackage{bm}
\usepackage{cite}
\usepackage{amsmath}
\usepackage{indentfirst} 
\usepackage{tikz}
\usepackage{longtable,multirow}
\usetikzlibrary{shapes,arrows}

\def\eg{{\it e.g.}}
\def\ie{{\it i.e.}}

\newcommand{\br}{\mbox{\ensuremath{\mathcal{B}}}}
\newcommand{\sigmaXBF}{\mbox{\ensuremath{\sigma\times\mathcal{B}}}\xspace}

\def\chitz{\ensuremath{\tilde{\chi}^0_2}}
\def\chiz{\ensuremath{\tilde{\chi}^0_1}}
\def\chipm{\ensuremath{\tilde{\chi}^\pm_1}}

\def\tchi{\ensuremath{\tilde{\chi}}}

\def\slep{\ensuremath{\tilde{l}}\xspace}

\def\spacer{\hspace*{10mm} }

\def\lsim{\mathrel{\raise.3ex\hbox{$<$\kern-.75em\lower1ex\hbox{$\sim$}}}}
\def\gsim{\mathrel{\raise.3ex\hbox{$>$\kern-.75em\lower1ex\hbox{$\sim$}}}}
\def\ifmath#1{\relax\ifmmode #1\else $#1$\fi}

\begin{document}
\begin{center}

\begin{flushright}
LPSC14001\\
HEPHY-PUB 932/13
\end{flushright}

\vspace*{1.6cm}
{\Large\bf SModelS: a tool for interpreting simplified-model results\\[2mm] from the LHC and its application to supersymmetry} 

\vspace*{1cm}\renewcommand{\thefootnote}{\fnsymbol{footnote}}

{\large 
Sabine~Kraml$^{1}$\footnote[1]{Email: sabine.kraml@lpsc.in2p3.fr},
Suchita~Kulkarni$^{1}$\footnote[2]{Email: suchita.kulkarni@lpsc.in2p3.fr},
Ursula~Laa$^{2}$\footnote[3]{Email: ursula.laa@assoc.oeaw.ac.at},
Andre~Lessa$^{3}$\footnote[4]{Email: lessa@if.usp.br},\\[1mm]
Wolfgang~Magerl$^{2}$\footnote[5]{Email:  wolfgang.magerl@assoc.oeaw.ac.at},
Doris~Proschofsky-Spindler$^{2}$\footnote[6]{Email: doris.proschofsky@assoc.oeaw.ac.at},
Wolfgang~Waltenberger$^{2}$\footnote[7]{Email: wolfgang.waltenberger@oeaw.ac.at}
} 

\renewcommand{\thefootnote}{\arabic{footnote}}

\vspace*{1cm} 
{\normalsize \it 
$^1\,$Laboratoire de Physique Subatomique et de Cosmologie, 
Universit\'e Grenoble-Alpes, CNRS/IN2P3, 53 Avenue des Martyrs, F-38026 Grenoble, France\\[2mm]
$^2\,$Institut f\"ur Hochenergiephysik,  \"Osterreichische Akademie der Wissenschaften,\\ Nikolsdorfer Gasse 18, 1050 Wien, Austria\\[2mm]
$^3\,$Instituto de F\'isica, Universidade de S\~ao Paulo, S\~ao Paulo - SP, Brazil\\[2mm]
}

\vspace{1cm}

\begin{abstract}
We present a general procedure to decompose Beyond the Standard Model (BSM) collider signatures 
presenting a $\mathbb{Z}_2$ symmetry into Simplified Model Spectrum (SMS) topologies. 
Our method provides a way to cast BSM predictions for the LHC in a model independent framework, 
which can be directly confronted with the relevant experimental constraints. 
Our concrete implementation currently focusses on supersymmetry searches with missing energy, 
for which a large variety of SMS results from ATLAS and CMS are available.  
As show-case examples we apply our procedure to two scans of the minimal supersymmetric standard model.  
We discuss how the SMS limits constrain various particle masses and which regions of parameter space 
remain unchallenged by the current SMS interpretations of the LHC results.  
\end{abstract}

\end{center}

\clearpage

\section{Introduction}\label{sec:intro}

Searches at the ATLAS and CMS experiments at the LHC show no signs of physics beyond the
Standard Model (BSM). 
After the first phase of LHC operation at centre-of-mass energies of 7--8~TeV in 2010--2012, 
the limits for the masses of supersymmetric particles, in particular of 1st/2nd generation
 squarks and gluinos, have been pushed well into the TeV range~\cite{atlas:susy:twiki,cms:susy:twiki}. 
Likewise, precision measurements in the flavor sector, in particular in $B$-physics, 
are well consistent with Standard Model (SM) expectations~\cite{Amhis:2012bh,lhcb:2012ct} 
and show no sign, or need, of new physics. 
At the same time the recent discovery~\cite{atlas:2012gk,cms:2012gu} of a Higgs-like particle 
with mass around 125~GeV makes the question of stability of the electroweak scale---the 
infamous gauge hierarchy problem---even  more imminent. 
Indeed, supersymmetry (SUSY) is arguably the best-motivated theory to solve the gauge 
hierarchy problem and to explain a light SM-like Higgs boson. 
So, the Higgs has very likely been discovered---but where is supersymmetry? 

Looking closely \cite{Sekmen:2011cz,Arbey:2011un,Papucci:2011wy,CahillRowley:2012kx,Dreiner:2012gx,Mahbubani:2012qq} 
one soon realizes that many of the current limits on SUSY particles are based 
on severe model assumptions, which impose particular relations between particle masses, decay 
branching ratios, {\it etc}. The prime example is the interpretation of the search results within the 
Constrained Minimal Supersymmetric Standard Model (CMSSM). 
The interpretation of the search results within a much more general  realization 
of the MSSM is perfectly feasible, see \cite{Sekmen:2011cz,Arbey:2011un,SUS-12-030}, 
but computationally very demanding and certainly not suitable for a ``quick'' survey. 

An approach which has therefore  
been adopted systematically by the ATLAS and CMS collaborations, 
is to interpret the results within so-called Simplified Model Spectra \cite{Okawa:2011xg,cms:2013wc}.  
Simplified Model Spectra, or SMS for short, are effective-Lagrangian descriptions involving  
just a small number of new particles. They were designed as a useful tool for the characterization 
of new physics, see \eg\ \cite{Alwall:2008ag,Alves:2011wf}. 
A large variety of results on searches in many different channels are available from both ATLAS and CMS, 
providing general cross section limits for SMS topologies. 
However, using these results to constrain complex SUSY (or general BSM) 
scenarios is not straightforward. 

In this paper, we present a method to decompose the signal of an arbitrary SUSY 
spectrum into simplified model  
topologies and test it against all the existing LHC bounds in the SMS context. 
The computer package doing all this is dubbed {\tt SModelS}~\cite{smodels}. 
(A similar approach was in fact proposed some time ago in \cite{ArkaniHamed:2007fw}.)
As we will show, this decomposition allows a vast survey of SUSY models, and enormously simplifies the task of identifying the regions of parameter space which are still allowed by the current searches.
Our method also allows us to discuss the current coverage of the simplified models considered so far
by the ATLAS and CMS searches. We can, for instance, identify possible regions 
of parameter space which are not tested by any of the simplified models assumed by the current searches.
Two scans of the MSSM with parameters defined at the weak scale, one with 7 and one with 9 free parameters,  
are used as show-case examples to demonstrate the use of our method. 
Our results show that the SMS interpretation of the LHC results indeed provide important constraints on SUSY scenarios. 
At the same time, however, large regions of interesting parameter space with SUSY particles below 1~TeV 
remain unchallenged by the current SMS results. 

It is important to note that, while our method was originally developed with SUSY searches in mind, 
and the application presented here focusses on the MSSM, 
our approach  is perfectly general and easily extendible to any BSM model to which the experimental SMS results apply. 

The rest of this paper is organized as follows. 
In Section~\ref{sec:overview} we briefly describe the general procedure for the decomposition and the use of the  experimental results. A more detailed description is given in Section~\ref{sec:details}; readers who are not interested in technicalities may skip this section altogether. 
Section~\ref{sec:validation}  discusses the validation of the {\tt SModelS} framework.  
In Section~\ref{sec:scan} we then apply {\tt SModelS} in two scans  of the MSSM and discuss how the SMS limits constrain various particle masses, and which regions of parameter space remain untested. Conclusions and an outlook are given in Section~\ref{sec:conclusions}.

\section{General Procedure}\label{sec:overview}

Currently most ATLAS and CMS experimental analyses consider specific simplified models to 
present their constraints on new physics. The number of signal events expected in a given signal 
region is then obtained using the signal efficiency times acceptance for the 
specific model assumed. In general, the signal efficiency is model dependent and must be 
calculated for each specific model considered. Nonetheless, most current experimental analyses 
aim for model independent constraints and consider sufficiently inclusive signal regions.
The guiding principle behind the procedure discussed here is the assumption that the signal efficiencies
for most experimental searches for new physics depend mostly on the event kinematics and are just marginally
affected by the specific details of the BSM model. 
This allows us to map the full model's signals to its SMS-equivalent topologies and use the latter 
to constrain the full model.

Clearly, this assumption, which from now on we call SMS assumption, is not always valid. 
For instance, it is expected to be violated for searches which strongly rely on shape distributions.
The signal efficiencies can moreover depend on spin correlations, or on properties of off-shell states 
in production channels (s-channel or t-channel production) or in decays. 
A quantitative discussion of the sensitivity of signal efficiencies to various model properties is 
analysis dependent and a complicated matter, being out of the scope of the current work. 
Dedicated studies will be presented in future publications. 
Generally, it is the responsibility of the user to apply {\tt SModelS} only to models and experimental results 
for which the SMS assumption is approximately valid. 
(Note that it is possible to use only a subset of experimental results in the database.) 
Let us now outline the general {\tt SModelS} procedure. 

\begin{figure}[t!]\centering
\includegraphics[width=0.95\textwidth]{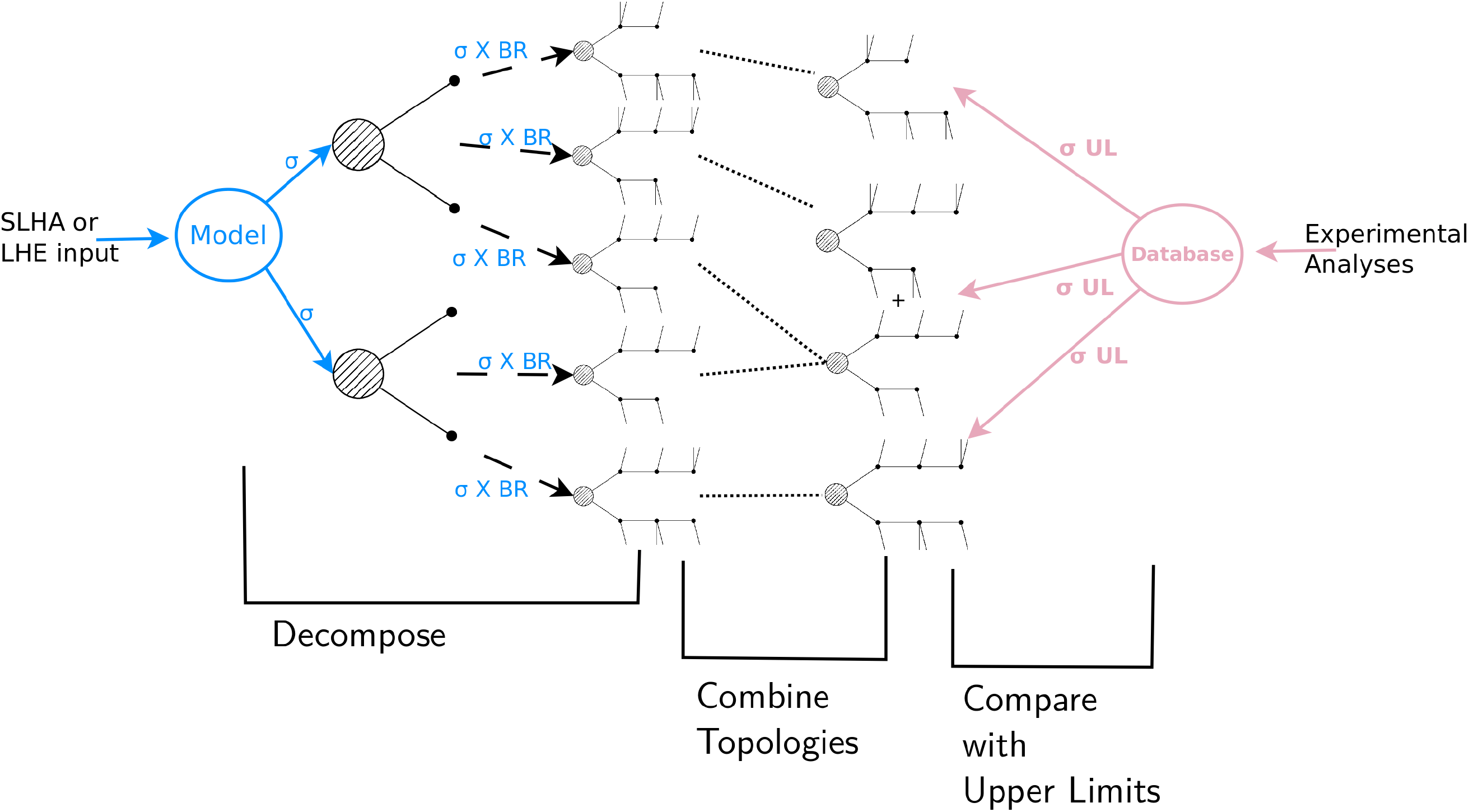}
\caption{\label{fig:scheme}
Schematic view of the working principle of {\tt SModelS}.}
\end{figure}

Under the SMS assumption, to a first approximation, it is possible to reduce all the properties of a 
BSM model to its mass
spectrum, production cross sections ($\sigma$) and decay branching ratios ($\mathcal{B}$).  
With this knowledge we can decompose the full BSM signal in a series of independent signal topologies with their specific weights given
by the corresponding production cross section times branching ratio (\sigmaXBF). Such a decomposition is extremely helpful to cast the theoretical
predictions of a specific BSM model in a model-independent framework, which can then be compared against 
the experimental limit on this \sigmaXBF. A schematic view of the working principle is given in Fig.~\ref{fig:scheme}.

\begin{figure}[t!]
\begin{center}
\includegraphics[width=0.35\linewidth]{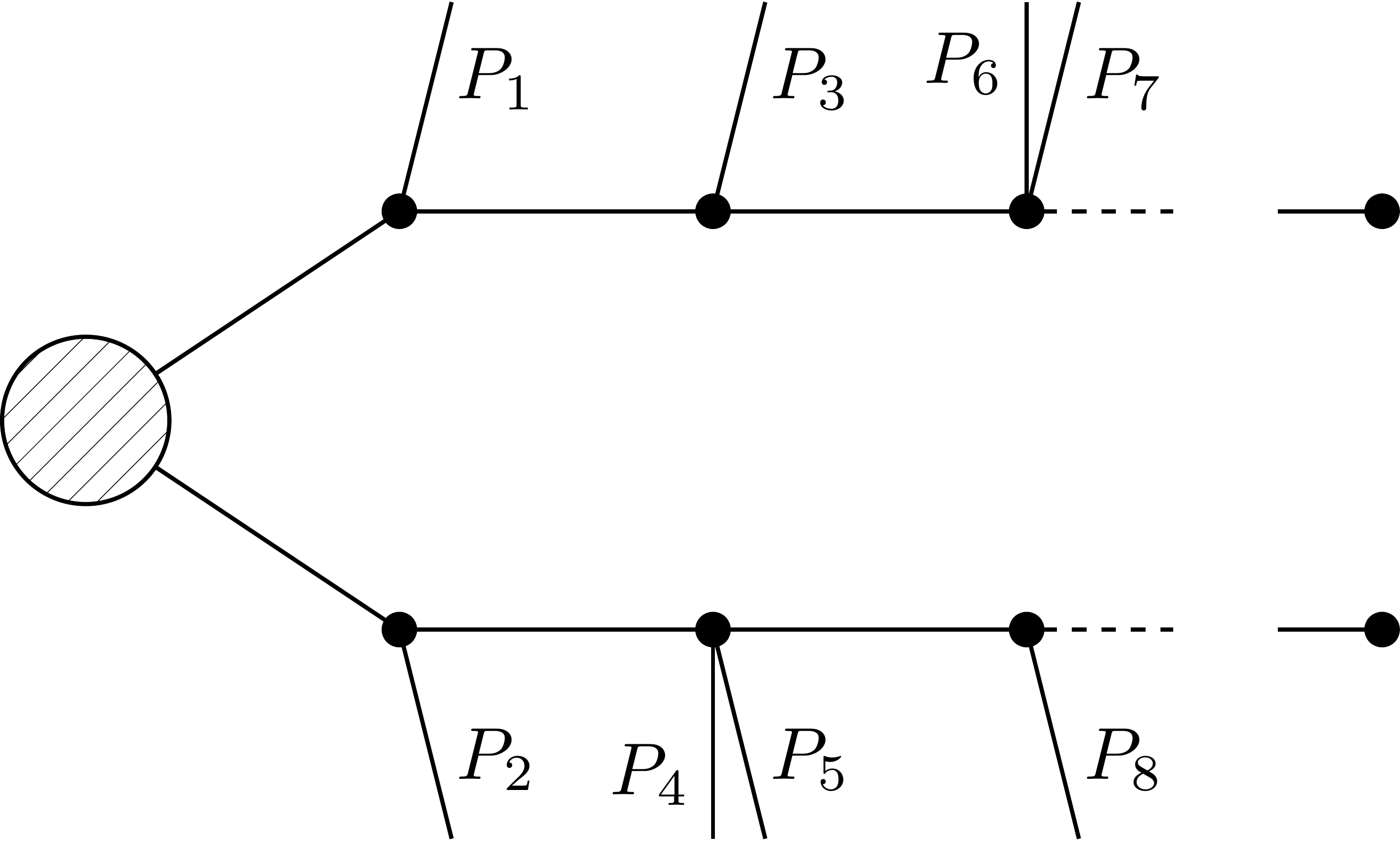}
\caption{The general type of SMS topology considered in this paper. The $P_i$ label the SM final state particles. 
The end of each decay chain is always the lightest $\mathbb{Z}_2$-odd particle which is stable.}
\label{fig:GenSMSTop}
\end{center}
\end{figure}

\begin{figure}[t!]
\begin{center}
\includegraphics[width=0.42\linewidth]{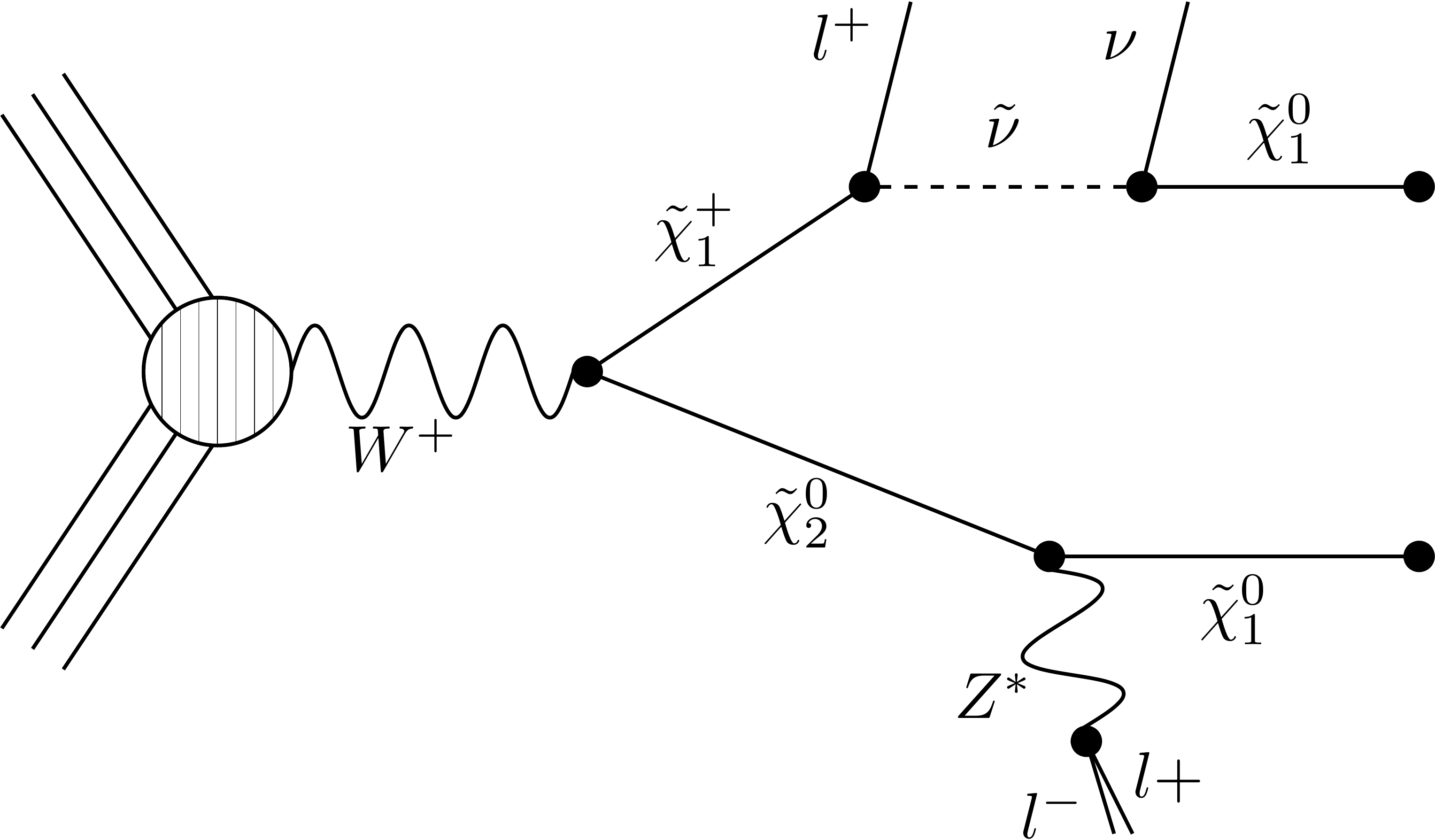}\spacer
\begin{picture}(40,100)(10,20)
\put(0,80){\vector(1,0){30}}
\end{picture}
\includegraphics[width=0.32\linewidth]{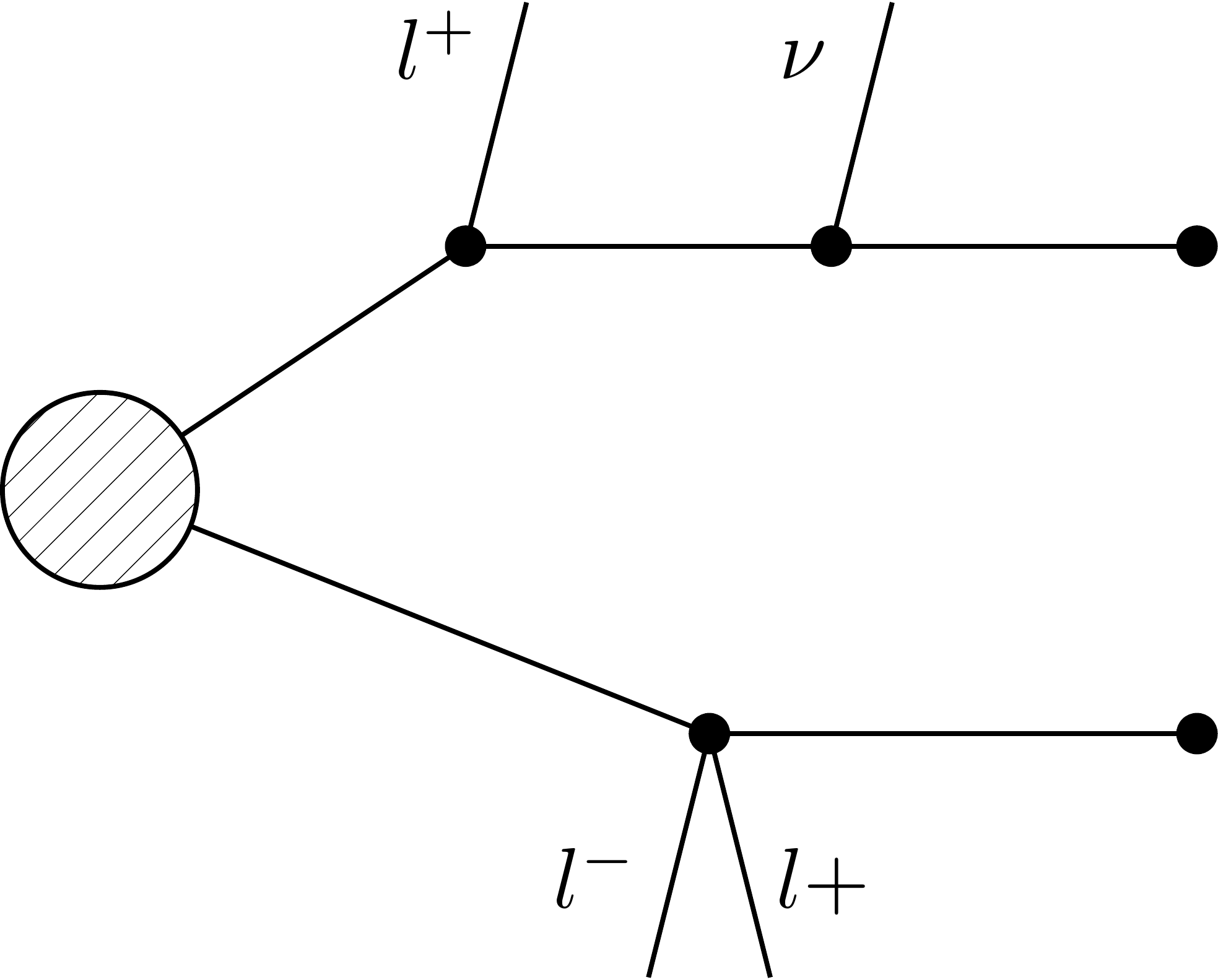}\spacer \\
\caption{A full model diagram (left) and its SMS equivalent topology (right).}
\label{fig:SMSexample}
\end{center}
\end{figure}
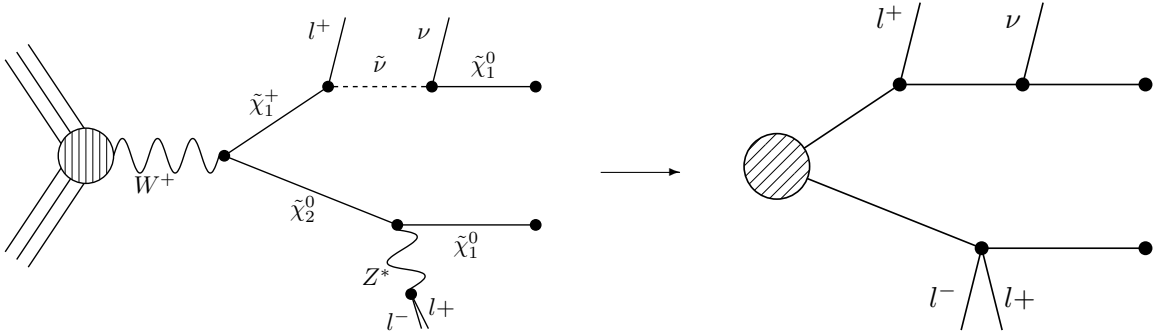

The first step in our procedure is to compute all signal topologies appearing in the full model\footnote{By full model we mean, for instance, a specific MSSM parameter point.} and their
respective weights, $\sigmaXBF$. Since here we only consider models with a $\mathbb{Z}_2$ symmetry 
(such as R-Parity, T-Parity or KK-Parity), the possible signal topologies will always arise 
from pair production of new $\mathbb{Z}_2$-odd particles,
which decay as $P \to P' + {\rm SM}$ particles,
where $P$ and $P'$ are the parent and daughter BSM particles, respectively.\footnote{Throughout this work 
we ignore BSM particles which are $\mathbb{Z}_2$ even, such as heavy Higgs bosons in the MSSM. 
The light MSSM Higgs $h$ is treated as a SM particle.} 
Hence all topologies will be of the form represented in Fig.~\ref{fig:GenSMSTop}, which shows the production
of the initial pair of BSM states (represented by a circle with two outgoing legs) and their subsequent cascade decays.
In our notation all particles appearing in the SMS topology,  
both $\mathbb{Z}_2$-even and $\mathbb{Z}_2$-odd, are on-shell. 
The case of off-shell decays is always included as 3-body decays, with no mention to the off-shell states.
Therefore all the relevant information (in the SMS framework) of such a diagram can be 
reduced to three main objects:
\begin{itemize}
\item the diagram topology: number of vertices and SM final state particles in each vertex;
\item the masses (mass vector) of the $\mathbb{Z}$-odd BSM particles appearing in the diagram;
\item the diagram weight (\sigmaXBF).
\end{itemize}
The reduction of a particular process to its equivalent SMS topology is
illustrated in Fig.~\ref{fig:SMSexample}.
Details of the decomposition procedure and the labeling scheme used are explained in Section~\ref{sec:decomposition}. 
Note that once the decomposition is done, the full model is reduced to its signal topologies 
and there is no longer any reference to the specific details of the model, 
except for the relevant $\mathbb{Z}_2$-odd masses and the \sigmaXBF associated to each topology. 
In this way we can cast the theoretical predictions in a model-independent way.\footnote{One has to keep 
in mind, however, that the color factor of the initially produced BSM particles
influences the QCD activity in the final state and may thus significantly affect the signal efficiency. 
This is not a worry in the following as we did not come across any example 
where constraints from a experimental result assuming QCD production are used
to exclude an EW produced topology, or vice-versa, but one might encounter such
cases in principle.}

The next and more involved step is to confront the theoretical predictions obtained from the decomposition
with the experimental constraints. For that it is
necessary to map the signal topologies produced in the decomposition to the SMS topologies
constrained by data.
For some experimental analyses this is a trivial matter, since they provide an upper limit for a single topology
cross section as a function of the relevant BSM mass vector. Examples are constraints on squark pair
production, with $\tilde q\to q+\chiz$, which give an upper limit on \sigmaXBF as a function of $(m_{\tilde{q}},m_{\chiz})$, or gluino pair production, with
$\tilde g\to t\bar t+\chiz$, which limit \sigmaXBF as a function of $(m_{\tilde{g}},m_{\chiz})$. 
However it is often the case that the experimental analysis does not constrain a single topology but rather a sum of several topologies, assuming a specific relative contribution from each of them. As an example, consider the slepton pair production limits,
where the interpretation constrains the sum over final state lepton flavors ($e$'s and $\mu$'s) under the assumption that
each flavor contributes 50\% to the signal and that selectrons and smuons are mass degenerate,  
$(m_{\tilde{e}},m_{\chiz}) = (m_{\tilde{\mu}},m_{\chiz})$. 
In order to apply this experimental constraint to the signal topologies obtained from the decomposition, it
is necessary to combine all topologies with a single lepton being emitted in each branch and which have the same mass vector.
Moreover, in order for the experimental constraint to be valid, it is necessary to verify the analysis conditions: 
topologies with $e$'s and $\mu$'s contribute equally to the final theoretical prediction (\sigmaXBF).
A more involved example are the constraints from tri-lepton+MET searches: here the SMS results for $\chipm\chitz$ production with decays through sleptons assume the intermediate slepton being a selectron, smuon  or stau (including or not including sneutrinos) so that the limits on \sigmaXBF only apply for specific flavor-democratic, tau-enriched, tau-dominated, etc., cases. Such constraints need to be carefully taken into account when mapping the signal topologies obtained from the decomposition to the experimental results. A detailed description
of how the analyses assumptions and constraints are described in a model independent language is presented in Sec.~\ref{sec:database}.
The procedure for matching the decomposition results to the analysis constraints is discussed in detail in Sec.~\ref{sec:interface}.

Finally when all signal topologies are combined according to the assumptions of each experimental result, 
the resulting theoretical predictions for the cross sections of the combined topologies can be directly compared
to the experimental upper limits.
Thus it can be decided whether a particular parameter point (a particular BSM spectrum) is excluded or not 
by the available SMS results.

\section{Detailed Description} \label{sec:details}

In this section, we discuss the technical details of the main building blocks of {\tt SModelS}: 
the decomposition procedure, the analysis database, and the matching of theoretical and 
experimental results.

\subsection{Decomposition Procedure}   \label{sec:decomposition}

As explained in the previous section, under the SMS assumption, all the complexity of the BSM model 
can be replaced by the knowledge of the signal topologies and their weights, together with the 
relevant BSM masses.\footnote{As discussed in Section \ref{sec:overview}, the SMS assumption may be violated
for specific analyses given a particular model. For instance, if the signal efficiency for a specific analysis 
is too sensitive to the shape of the signal distributions and these are strongly affected by the type of production channel
(s-channel or t-channel) or by the nature of the off-shell states mediating the decays, the reduction of a full model
to its SMS topologies no longer encapsulates all the necessary information.
We leave it to the user to evaluate which analyses may be applied for the input model considered.}  
The topologies can then be classified according to the 
number of vertices in each branch and the SM states appearing in each vertex.
In order to properly classify the signal topologies, we introduce a formal labeling scheme, which is
{\em i)} model independent,
{\em ii)} general enough to describe any topology,
{\em iii)} sufficiently concise and
{\em iv)} allows us to easily combine topologies according to the assumptions and conditions in the experimental analyses.

\begin{figure}[h!]
\begin{center}
\includegraphics[width=0.65\linewidth]{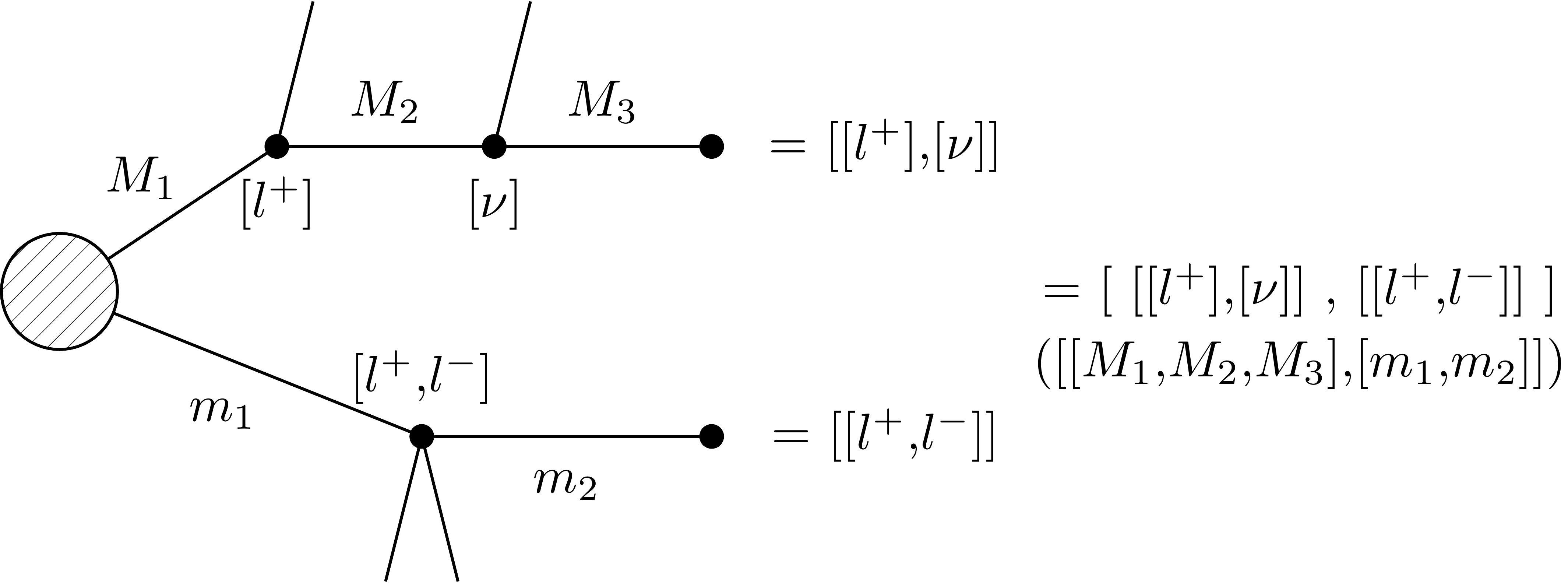}
\caption{The labeling scheme adopted here applied to an example diagram. In parenthesis we show the respective mass vector for the diagram.}
\label{fig:SMSlabel}
\end{center}
\end{figure}

We choose to use a labeling system based on nested brackets, which corresponds to a textual representation of the topology.
As discussed in Section~\ref{sec:overview}, we assume that all signal topologies respect a $\mathbb{Z}_2$-symmetry,
hence BSM states are produced in pairs and cascade decay to a single BSM state and SM particles.
Therefore a signal topology always contains two branches (one for each of the initially pair produced BSM states),
which we describe by $[B_1,B_2]$.
In each branch there is a series of vertices, which represent the cascade decays. From each vertex there is one outgoing BSM
state and a number of outgoing SM particles.
Thus, inside each branch bracket, we add an inner bracket for each vertex, containing the lists of outgoing SM particles.
Note that there is no mention to the intermediate BSM particles ($\tchi_1$, $\slep$,...),
which makes our method explicitly model independent. The only information kept from the BSM states
are their masses.
To illustrate the labeling scheme just defined, we show in Fig.~\ref{fig:SMSlabel} a signal topology containing a 2-step
cascade decay in one branch and a one step decay in the other. Following the above prescription, this topology is
described by $[B_1,B_2]$, with $B_1 = [[\mbox{particles in vertex 1}],[\mbox{particles in vertex 2}]]$ and $B_2=[[\mbox{particles in vertex 1}]]$, as seen in Fig.~\ref{fig:SMSlabel}. Note that the brackets inside each branch are ordered according to the vertex number.
For the specific final SM particles assumed in the figure, 
we finally obtain: topology $= [\;[[l^+],[\nu]]\;,\;[[l^+,l^-]]\;]$. 
The corresponding mass vector, shown in parenthesis in Fig.~\ref{fig:SMSlabel}, is given by $[[M_1,M_2,M_3],[m_1,m_2]]$, 
where once again the masses are ordered according to the vertices. Note that we allow for the possibility of distinct final state masses
in each branch ($M_3 \neq m_2$). Finally, adding the topology weight ($\sigmaXBF$), 
we have a full description of the signal topology.

Given a BSM model ---described by its spectrum, branching ratios and production cross sections--- we need to obtain
all the possible signal topologies and compute their weights ($\sigmaXBF$).
The procedure of computing these corresponds to the decomposition of the full model in terms of simplified model topologies.
There are two ways to actually perform the SMS decomposition. The first requires the generation of parton-level Monte Carlo (MC) events,
followed by the mapping of each event into the corresponding SMS topology. 
The second method is purely based on the SLHA~\cite{Skands:2003cj} spectrum file and decay table, supplemented by theoretically computed cross sections.  Below we describe both approaches in more detail.

\subsubsection*{Monte Carlo based decomposition }

The decomposition based on parton level MC events is the most general one, since it can decompose any type of BSM model,
as long as it is possible to simulate MC events for it. In this case an event file in the LHE format is used as input and 
each event is mapped to a simplified model topology. The mapping is obviously not one-to-one, since more than one event can
generate the same exact SMS topology. Then the sum of MC weights for all events contributing to the same topology
directly gives $\sigmaXBF$ for the corresponding topology. The disadvantage of this method is the introduction of MC uncertainties in the decomposition result; this can however be easily solved by increasing the MC statistics.
We also note that the recent advances on NLO MC generators will allow to produce decomposition results at NLO even in the MC based decomposition.

\subsubsection*{SLHA-based decomposition }

In the SUSY case, there are a number of public codes for the computation of the mass
spectrum and decay branching ratios with output in the SLHA~\cite{Skands:2003cj} format. 
The production cross sections can then be computed at leading order (LO) through a Monte Carlo generator, 
or at  next-to-leading order (NLO) with Prospino~\cite{Beenakker:1996ed}.
For gluinos and squarks, cross sections at next-to-leading log (NLL) precision can be computed using NLL-Fast~\cite{nllfast}.
These cross sections can be included in the SLHA file~\cite{slha-xsext}, which then holds all the required information for the SMS decomposition.

In the SLHA-based decomposition the cross sections for pair production of BSM states and the corresponding branching ratios are used to
generate all possible signal topologies. The only theoretical uncertainties in this case
come from the cross-section uncertainty, hence being much smaller than the uncertainties in the MC-based method,
which depends on the MC statistics. In order to avoid dealing with a large number of irrelevant signal topologies, 
only the ones with $\sigmaXBF$ above a minimal cut value are kept.
For the results presented below we take this cut value to be $0.03$~fb.\footnote{Note that since experimental
results often constrain sums of topologies, individual signal topologies with small $\sigmaXBF$ may still be relevant for the final
theoretical prediction. This is why we take such a small value for the minimal cross section cut.}

\subsubsection*{Compression of topologies}

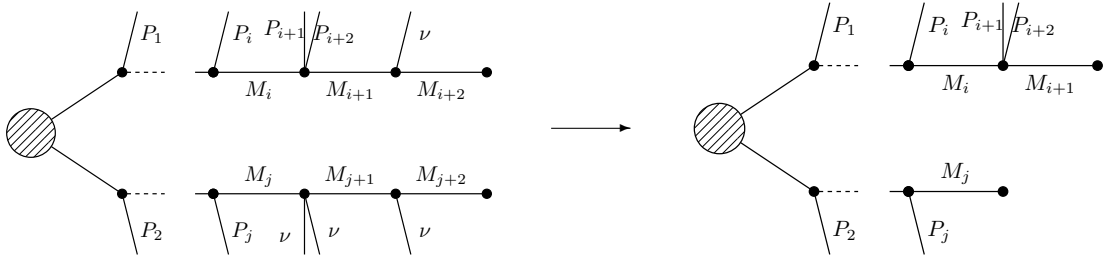
\begin{figure}[t!]
\begin{center}
\includegraphics[width=0.38\linewidth]{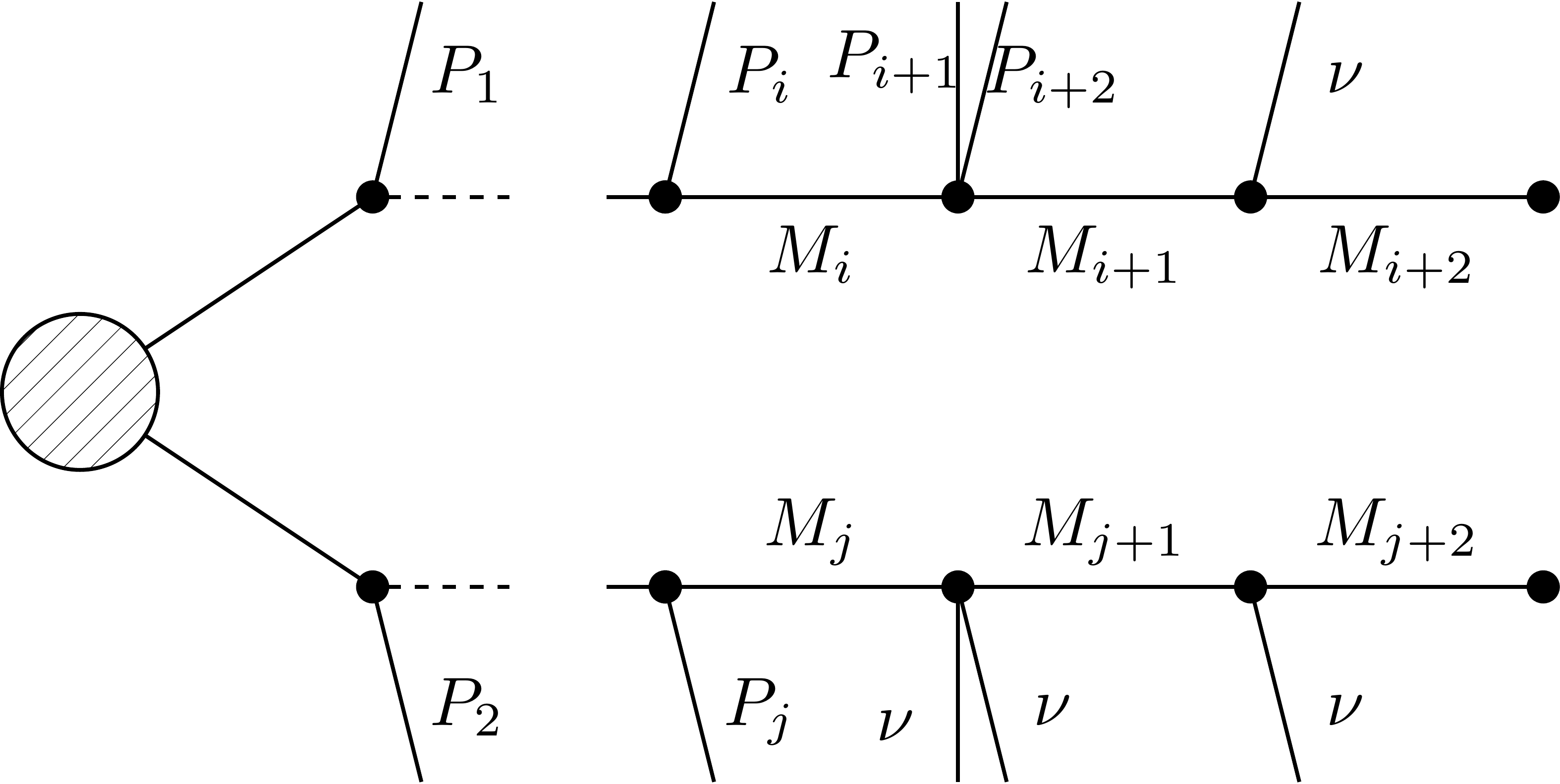}\spacer
\begin{picture}(40,100)(10,20)
\put(0,68){\vector(1,0){30}}
\end{picture}
\includegraphics[width=0.32\linewidth]{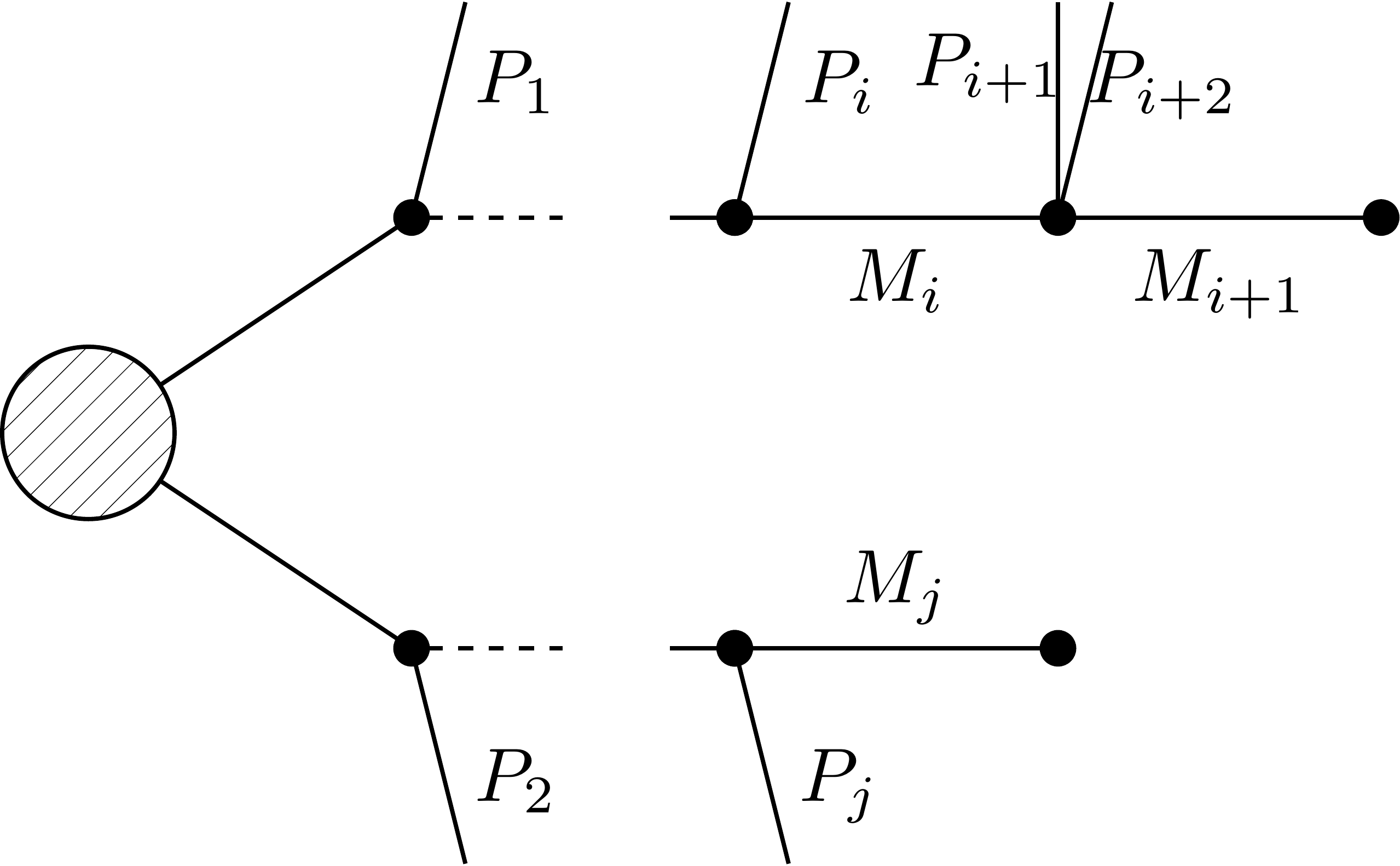}\spacer \\
\caption{A signal topology with invisible decays (left) and its [invisibly] compressed equivalent (right).
$M_{k}$ labels the masses of the BSM states appearing in the topology.}
\label{fig:Icompression}
\end{center}
\end{figure}

Although the decomposition described here (both for the MC and SLHA based methods) is fairly straightforward, 
the model independent language introduced above can be used to perform
non-trivial operations on the signal topologies. One possibility is the compression of topologies containing a series of invisible decays at the end of the cascade decay chain, as illustrated by Fig.~\ref{fig:Icompression}(left). 
In this case the effective final state BSM momentum is given by the
sum of the neutrinos and the final BSM state momenta. Therefore, for the experimental analyses, this topology
is equivalent to a compressed one, where the effective BSM final state includes the neutrino emissions, as
shown in Fig.~\ref{fig:Icompression}(right). Using the notation introduced previously,
this compression simply corresponds to $[[X',[\nu]],[X,[\nu,\nu],[\nu]]] \rightarrow [[X'],[X]]$, where $X,X'$ represent the other cascade decay
vertices. When performing this compression we must also remove the corresponding BSM masses from the mass vector,
so the effective final state BSM mass becomes heavier.
The compression procedure allows us to constrain the original topology using
experimental constraints for $[[X],[X']]$. More importantly, such ``invisible compression'' can be automatically
performed in a model-independent way using the nested bracket notation.

\begin{figure}[t!]
\begin{center}
\includegraphics[width=0.38\linewidth]{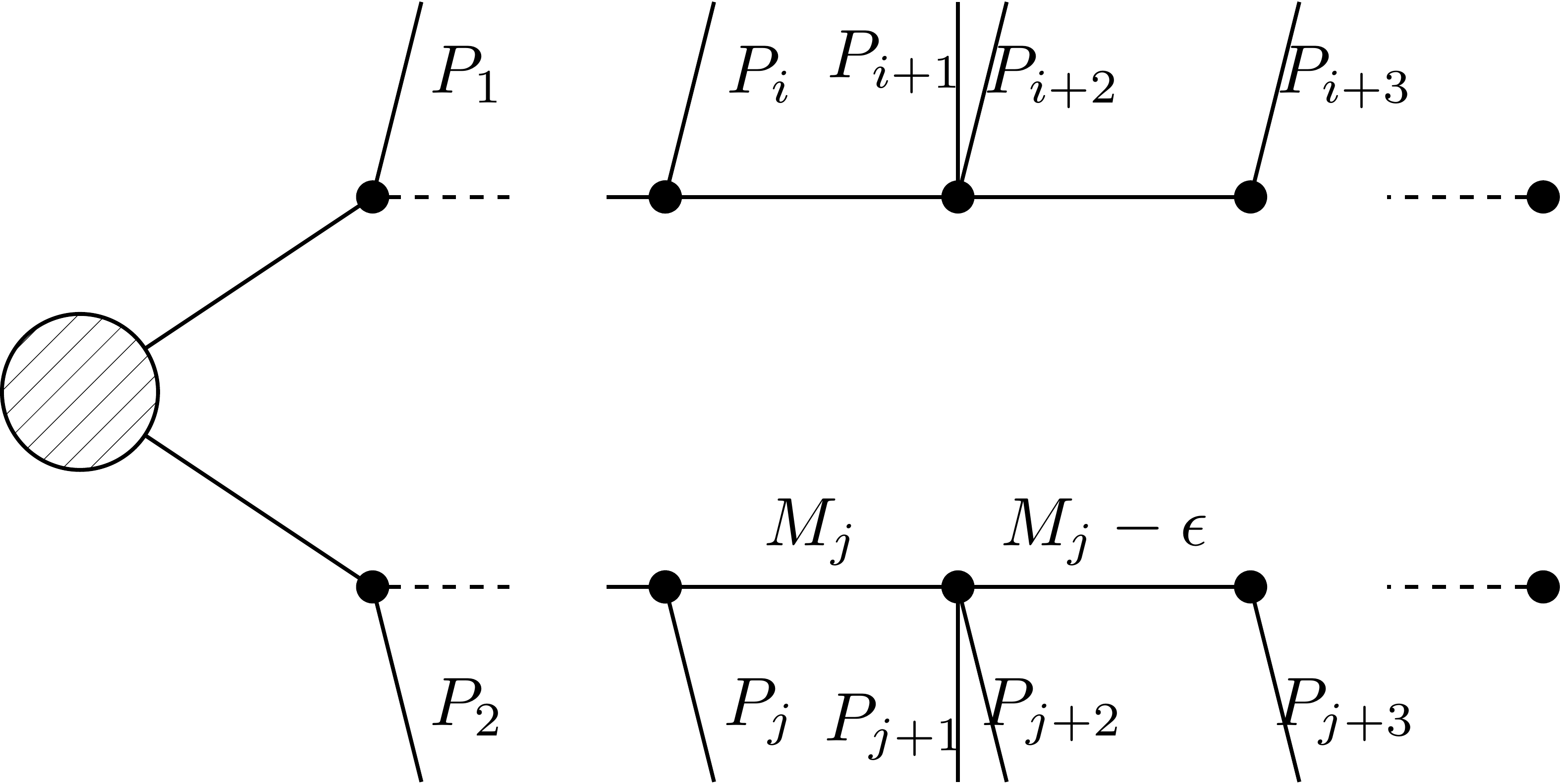}\spacer
\begin{picture}(38,100)(10,20)
\put(0,68){\vector(1,0){30}}
\end{picture}
\includegraphics[width=0.38\linewidth]{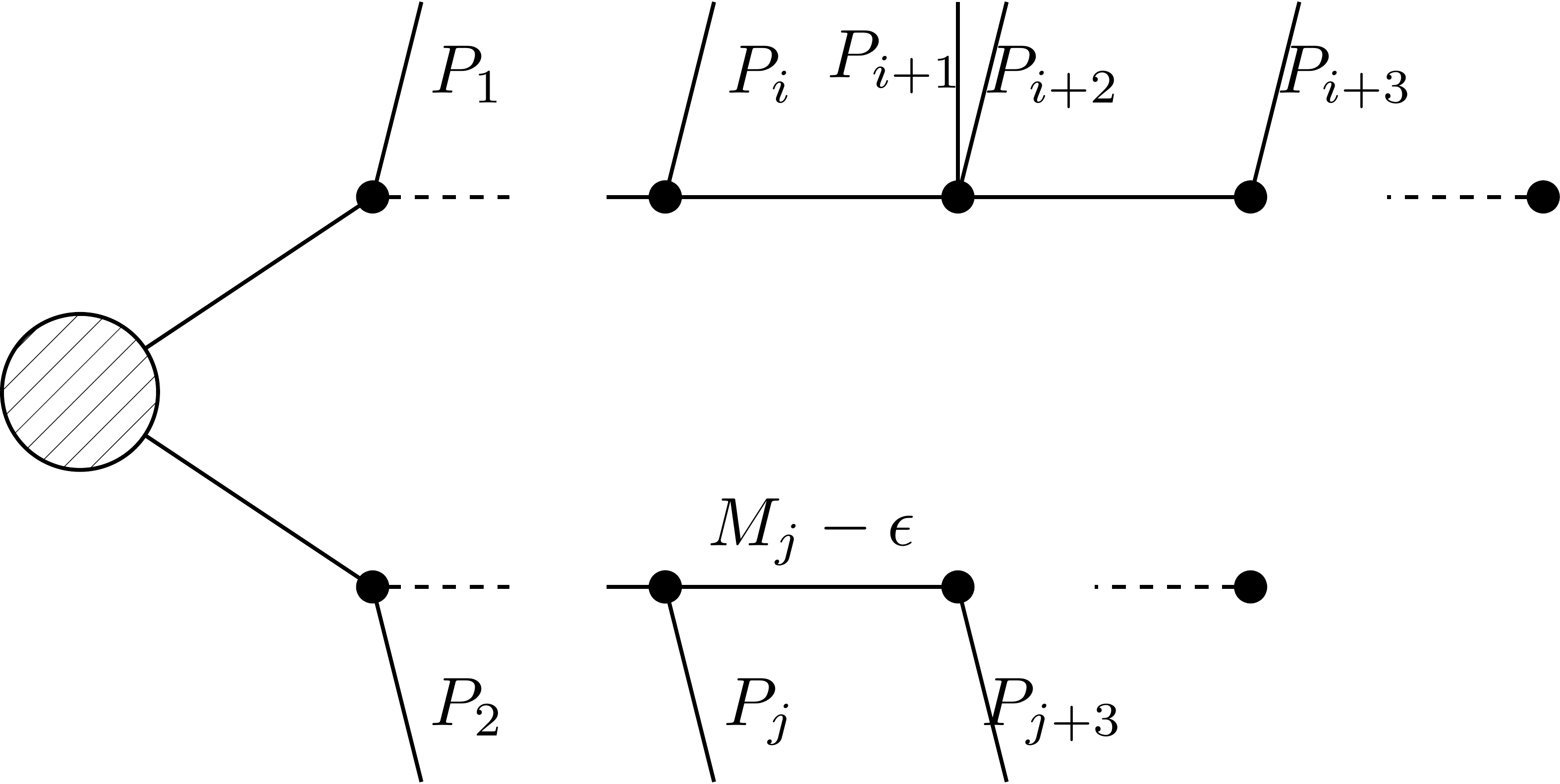}\spacer \\
\caption{A signal topology with a compressed mass spectrum (left) and its [mass] compressed equivalent (right).
$M_{j}$ and $M_{j}-\epsilon$ represent the masses of the quasi-degenerate states.}
\label{fig:Mcompression}
\end{center}
\end{figure}
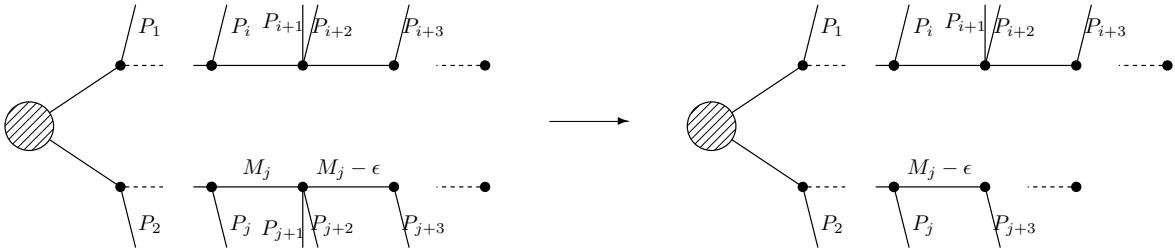

Another case of interest corresponds to topologies which contain spectra with very small mass splittings. 
In this case one can again approximately map the original topology into a smaller one,
where decays between quasi-degenerate states are omitted.
This ``mass compression'' approximation
is only reliable when the energy carried away by the SM particles emitted in the decay of quasi-degenerate
states is negligible for all experimental purposes.
If this is the case, the original topology is equivalent to a compressed one,
as shown in Fig.~\ref{fig:Mcompression}. For the results presented in Section~\ref{sec:scan} we
perform the mass compression for quasi-degenerate states if their mass difference is below 5 GeV.
Once again, this compression procedure 
allows us to constrain the original topology using experimental constraints for shorter cascade decays.
We also point out that, whenever (invisible and/or mass) compressions  are performed, care must be taken 
to avoid double counting topologies.
If both the original and the compressed topologies are kept after the decomposition, they should
not be combined later, since this would result in a double counting of the topology weight.

\subsection{Analysis Database}
\label{sec:database}

\subsubsection{Anatomy of an SMS result}

The interpretation of the BSM search results in the context of simplified models has become
the de-facto standard for the experimental collaborations. 
The ATLAS and CMS collaborations typically produce two types of SMS results: 
for each simplified model, values for the product of the experimental acceptance
and efficiency ($A \times\epsilon$) are determined to translate a number of signal events 
after cuts into a signal cross section. From this information, a 95\% confidence level upper limit (UL) 
on the product of the cross section and branching fraction (\sigmaXBF) is derived as a function 
of the BSM masses appearing in the SMS. 
Finally, assuming a theoretical ``reference'' cross section for each mass combination,
an exclusion curve in the plane of two masses is produced. 

\begin{figure}[t!] \centering
\includegraphics[width=0.52\linewidth]{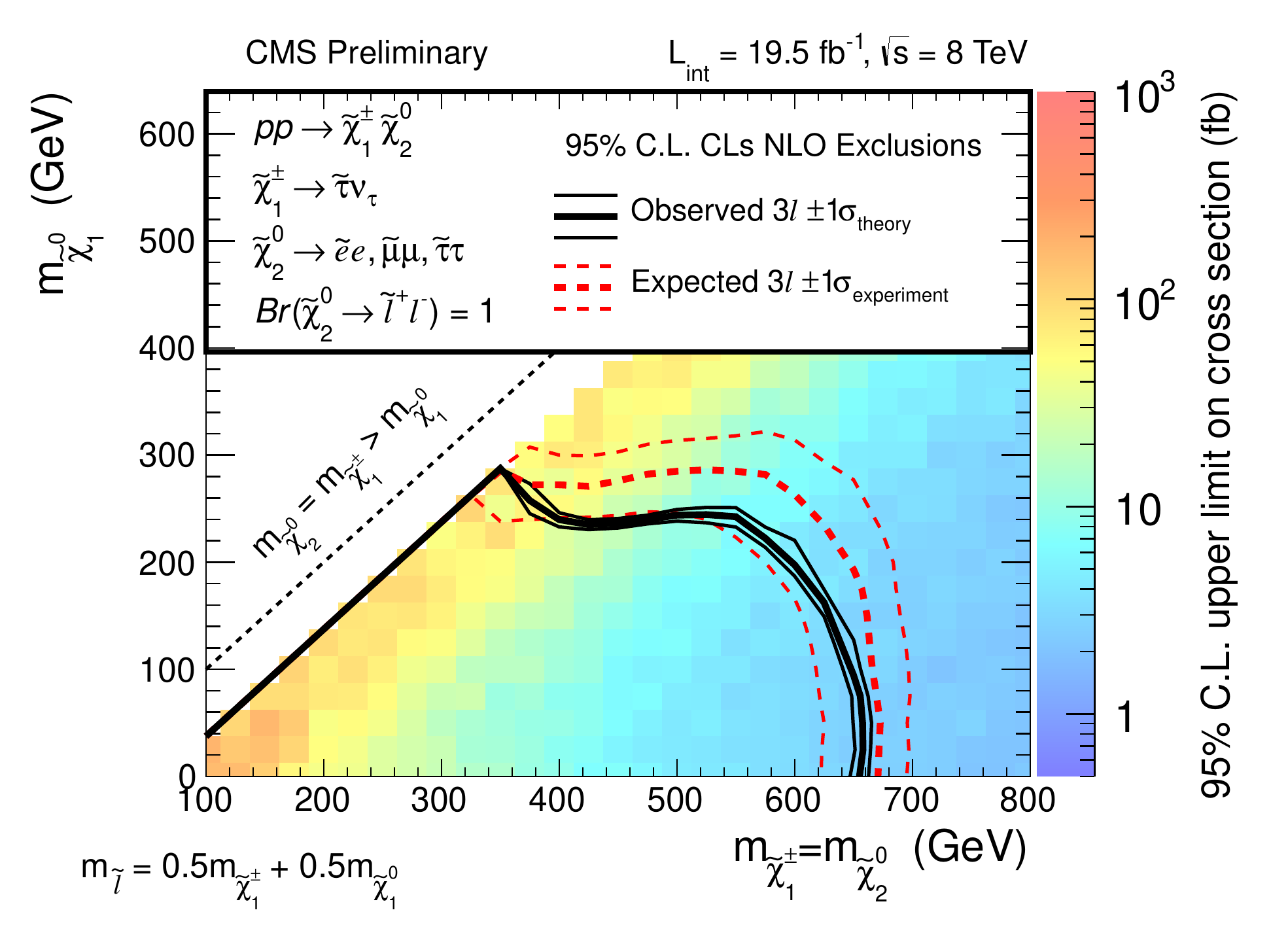}\quad
\includegraphics[width=0.43\linewidth]{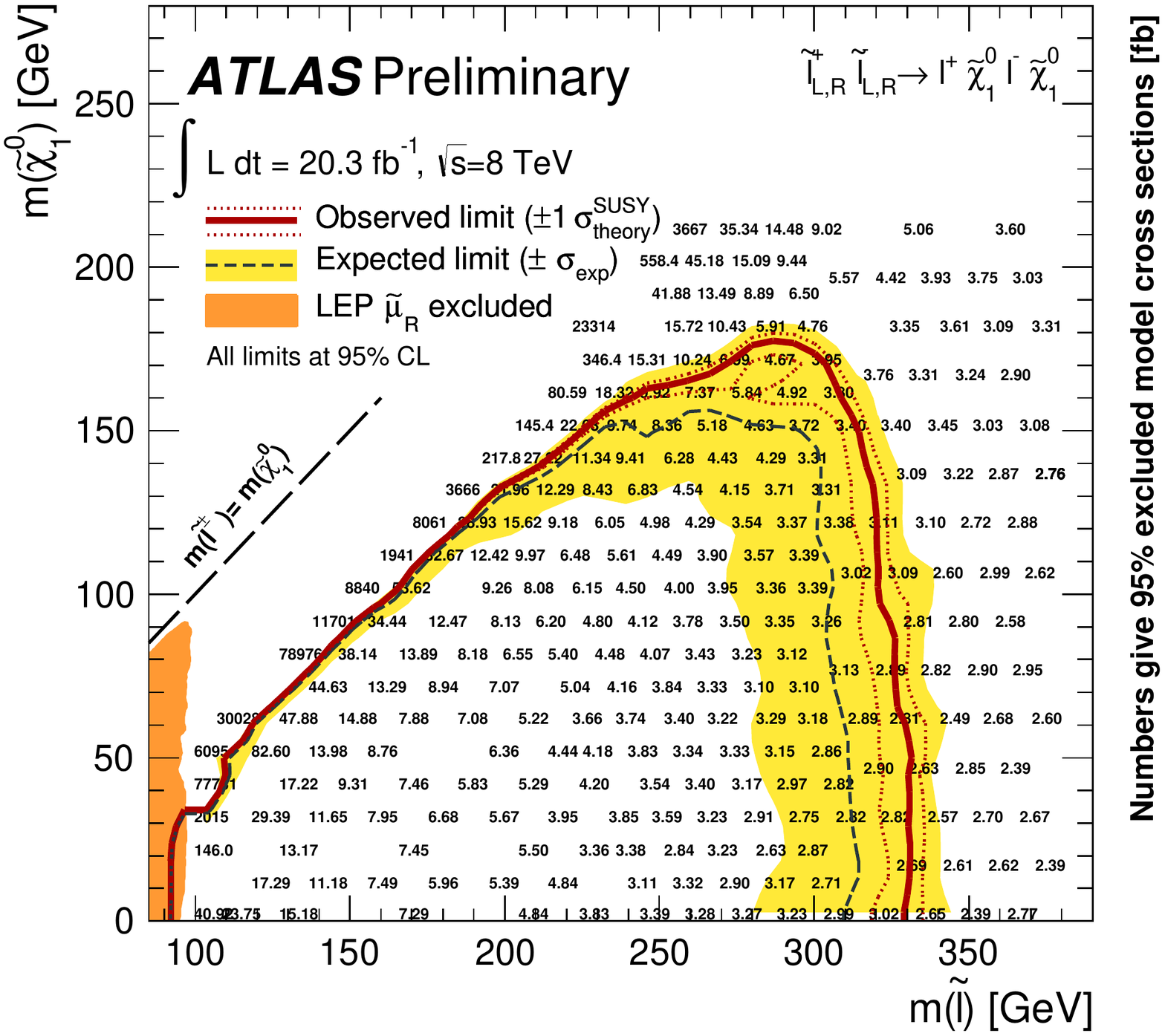}
\caption{Examples of SMS results from the CMS (left) and ATLAS (right)
collaborations, taken from~\cite{SUS-13-006} and \cite{ATLAS-CONF-2013-049}, respectively.}
\label{fig:smsexample}
\end{figure}

Figure~\ref{fig:smsexample} gives two examples. The plot on the left is from the CMS analysis  
of chargino--neutralino production with $\tilde\chi^{\pm}_1 \to \tilde \tau^{\pm}
\nu$ and $\tilde\chi^0_2 \to \tilde L^{\pm} L^{\mp}$, with $\tilde L^{\pm} \to
L^{\pm}\tilde\chi_1^0$, where $L = e,\mu,\tau$. Shown are the 95\% CL UL on \sigmaXBF together 
with the expected and observed mass limit curves. 
The plot on the right is an ATLAS result for the case of slepton pair production and 
direct decay to the LSP, $\tilde{l} \to l \tilde\chi^0_1$. 
Our approach builds upon the binned UL on \sigmaXBF, which is the information 
collected in our analysis database. 
Neither the efficiency plots nor the exclusion lines are used in our procedure.

The cross section upper limits reported in this way are however subject to a
number of assumptions made in the analysis, which also have to be described in the database.
In the {\tt SModelS} language described in Section~\ref{sec:decomposition}, the
topologies appearing in Fig.~\ref{fig:smsexample}\,(left) are $[[[L],[L]],[[\nu],[\tau]]]$, where $L =
e,\mu,\tau$ and the relevant mass vectors are $(m_{\tilde \chi_1^{\pm}},\, m_{\tilde L},\,
m_{\tilde\chi^0_1})$ and $(m_{\tilde \chi_2^0},\, m_{\tilde L},\,
m_{\tilde\chi^0_1})$.
The experimental analysis in our example assumes degenerate sleptons and
$m_{\tilde \chi_2^0} = m_{\tilde \chi_1^{\pm}}$,
so there are only three independent masses. Furthermore this
analysis constraints the sum over lepton flavors and charges for the
topologies listed above under the assumption that each term contributes
equally (flavor democratic decays).
Moreover,  the SMS result illustrated in Fig.~\ref{fig:smsexample}\,(left) is for the particular mass relation 
$m_{\tilde l}= (m_{\tilde\chi^{\pm}_1} + m_{\tilde\chi_1^0})/2$ and can
thus only be applied for (approximately) these $m_{\tilde l}$ and $m_{\tilde \nu}$ values.
However, in Refs.~\cite{SUS-13-006,ATLAS-CONF-2013-049}, results for distinct slepton mass 
values are also provided, which allows us to interpolate between them and apply these
results to more generic models.
All these constraints and conditions are described in our analysis database in the form of ``metadata''. 
As an example we show the metadata for the CMS result in Fig.~\ref{fig:smsexample} in
Table~\ref{tab:metadata1}. The entry {\it 'constraint'} lists the (sum of)
topologies being constrained by the analysis. If charges and/or flavors are not
explicitly listed, a sum over charges and/or flavors is assumed. The additional
analysis assumptions (each lepton flavor contributes equally to the total
weight, \sigmaXBF) can actually be relaxed because the efficiency for $\mu$'s
is higher than the efficiency for $e$'s, which is higher than the one for $\tau$'s. Therefore, instead of
requiring an equal weight (\sigmaXBF) from each flavor, it suffices to impose:
\[
\sigma\times\mathcal{B}\left(\tilde{l} \tilde{l} \to \mu^+\mu^-
\tilde\chi_1^0 \tilde\chi_1^0 \right) \geq
\sigma\times\mathcal{B}\left(\tilde{l} \tilde{l} \to e^+ e^-
\tilde\chi_1^0 \tilde\chi_1^0 \right) \geq
\sigma\times\mathcal{B}\left(\tilde{l} \tilde{l} \to \tau^+ \tau^-
\tilde\chi_1^0 \tilde\chi_1^0)\right).
 \]
These assumptions are included in the entry {\it 'fuzzycondition'} and the function
$Cgtr(x,y)$. This function uses the theoretical predictions for $x$ and $y$ to
map the condition $x > y$ into a number in the interval $0-1$.\footnote{The
actual function used is $Cgtr(x,y) = \left(|x-y| -
(x-y)\right)/\left(2(x+y)\right)$.} If $Cgtr(x,y) = 0$, the condition is fully
satisfied ($x > y$) and if $Cgtr(x,y) = 1$, the condition is fully violated ($x \ll y$). If the conditions are strongly violated ($Cgtr(x,y) > $ minimal value) by the input model, the corresponding analysis should not be used to constrain the model.
In this way,  the user can decide how strictly the conditions are enforced and ignore all constraints which
have a too large value of $Cgtr(x,y)$. In the following we ignore all constraints which have $Cgtr(x,y) > 0.2$.

\begin{table}[t!]
\caption{Metadata describing the SMS result from CMS in Fig.~\ref{fig:smsexample} (left plot).} 
\begin{verbatim}
sqrts: 8.00
lumi: 19.50
url: https://twiki.cern.ch/twiki/bin/view/CMSPublic/PhysicsResultsSUS13006
constraint: TChiChipmSlepStau -> [[['L'],['L']],[['nu'],['ta']]]
fuzzycondition: TChiChipmSlepStau ->
Cgtr([[['mu'],['mu']],[['nu'],['ta']]] , [[['e'],['e']],[['nu'],['ta']]])
Cgtr([[['e'],['e']],[['nu'],['ta']]] , [[['ta'],['ta']],[['nu'],['ta']]])
category: TChiChipmSlepStau -> eweakino
axes: TChiChipmSlepStau: M1 M0 005 - M1 M0 050 - M1 M0 095
\end{verbatim} 
\label{tab:metadata1}
\end{table}

Finally, the entry {\it 'axes'} describes the available slices of the
$(m_{\tilde \chi_1^{\pm}},\, m_{\tilde L},\,m_{\tilde
\chi_1^0})$ parameter space, which in this example corresponds to 
$m_{\tilde l}= 0.5m_{\tilde\chi^{\pm}_1} + 0.5m_{\tilde
\chi_1^0}$, $m_{\tilde l}= 0.95m_{\tilde\chi^{\pm}_1} + 0.05m_{\tilde
\chi_1^0}$ and $m_{\tilde l}= 0.05m_{\tilde\chi^{\pm}_1} + 0.95m_{\tilde
\chi_1^0}$. This information is used to interpolate between the values
 in the experimental results. Since different experimental analyses adopt
 distinct slicing methods of the parameter space, we use a general
 interpolation procedure which works for any slicing choice. It is based on
 a tesselation of the mass vector space and a linear interpolation on each
 simplex.
 
Another illustrative example is the ATLAS dilepton
search~\cite{ATLAS-CONF-2013-049} for slepton pair production and decay shown
in Fig.~\ref{fig:smsexample}\,(right):
$\slep^+ + \slep^- \to (l^+ \chiz) + (l^- \chiz)$, where $\slep = \tilde{e}, \tilde{\mu}$.
In this case selectrons and smuons are assumed to be mass degenerate and
the experimental collaboration constrains the sum of lepton flavors: $\sigma([[[e^+]],[[e^-]]]) + \sigma([[[\mu^+]],[[\mu^-]]])$.
Once again, each lepton flavor is assumed to contribute equally.
The metadata for this example is given in Table~\ref{tab:metadata2}. As in the
previous example, we assume the muon efficiency to be higher than the electron
one, hence in {\it 'fuzzycondition'} we require $\sigma([[[e^+]],[[e^-]]]) \leq
\sigma([[[\mu^+]],[[\mu^-]]])$ instead of equal flavor contributions.

\begin{table}[t!]
\caption{Metadata describing the SMS result from ATLAS in Fig.~\ref{fig:smsexample} (right plot).} 
\begin{verbatim}
sqrts: 8.00
lumi: 20.30
url: https://atlas.web.cern.ch/Atlas/GROUPS/PHYSICS/CONFNOTES/ATLAS-CONF-2013-049/
constraint: TSlepSlep -> [[['e+']],[['e-']]]+[[['mu+']],[['mu-']]]
fuzzycondition: TSlepSlep -> Cgtr([[['mu+']],[['mu-']]],[[['e+']],[['e-']]])
category: TSlepSlep -> directslep
axes: TSlepSlep: M1 M0
\end{verbatim}
\label{tab:metadata2}
\end{table}

\FloatBarrier

\subsubsection{List of analyses in database}

The analyses that are currently implemented in the database are:

\subsubsection*{Gluino and squark searches}
\begin{itemize}
\item {\bf ATLAS:} SUSY-2013-04~\cite{SUSY-2013-04},
ATLAS-CONF-2012-105~\cite{ATLAS-CONF-2012-105}, 
ATLAS-CONF-2013-007~\cite{ATLAS-CONF-2013-007}, 
ATLAS-CONF-2013-047~\cite{ATLAS-CONF-2013-047},
ATLAS-CONF-2013-061~\cite{ATLAS-CONF-2013-061},
ATLAS-CONF-2013-062~\cite{ATLAS-CONF-2013-062},
ATLAS-CONF-2013-089~\cite{ATLAS-CONF-2013-089}
\item {\bf CMS:} 
SUS-11-022~\cite{SUS-11-022}, 
SUS-11-024~\cite{SUS-11-024}, 
SUS-12-005~\cite{SUS-12-005}, 
SUS-12-011~\cite{SUS-12-011},
SUS-12-024~\cite{SUS-12-024},
SUS-12-026~\cite{SUS-12-026},
SUS-12-028~\cite{SUS-12-028}, 
SUS-13-002~\cite{SUS-13-002}, 
SUS-13-004~\cite{SUS-13-004}, 
SUS-13-007~\cite{SUS-13-007},
SUS-13-008~\cite{SUS-13-008}, 
SUS-13-012~\cite{SUS-13-012},
SUS-13-013~\cite{SUS-13-013}
\end{itemize}

\subsubsection*{Electroweakino searches}
\begin{itemize}
\item {\bf ATLAS:} 
ATLAS-CONF-2013-028~\cite{ATLAS-CONF-2013-028},
ATLAS-CONF-2013-035~\cite{ATLAS-CONF-2013-035},
ATLAS-CONF-2013-036~\cite{ATLAS-CONF-2013-036},
ATLAS-CONF-2013-093~\cite{ATLAS-CONF-2013-093}
\item {\bf CMS:} 
SUS-11-013~\cite{SUS-11-013}, 
SUS-12-006~\cite{SUS-12-006}, 
SUS-12-022~\cite{SUS-12-022},
SUS-13-006~\cite{SUS-13-006}, 
SUS-13-017~\cite{SUS-13-017}  
\end{itemize}

\subsubsection*{Direct slepton searches}
\begin{itemize}
\item {\bf ATLAS:}  ATLAS-CONF-2013-049~\cite{ATLAS-CONF-2013-049}
\item {\bf CMS:}  SUS-12-022~\cite{SUS-12-022}, SUS-13-006~\cite{SUS-13-006}
\end{itemize}

\subsubsection*{3rd generation: stop and bottom searches}
\begin{itemize}
\item {\bf ATLAS:} 
ATLAS-CONF-2012-166~\cite{ATLAS-CONF-2012-166},
ATLAS-CONF-2013-001~\cite{ATLAS-CONF-2013-001},
ATLAS-CONF-2013-007~\cite{ATLAS-CONF-2013-007},
ATLAS-CONF-2013-024~\cite{ATLAS-CONF-2013-024},
ATLAS-CONF-2013-025~\cite{ATLAS-CONF-2013-025},
ATLAS-CONF-2013-037~\cite{ATLAS-CONF-2013-037},
ATLAS-CONF-2013-047~\cite{ATLAS-CONF-2013-047},
ATLAS-CONF-2013-048~\cite{ATLAS-CONF-2013-048},
ATLAS-CONF-2013-053~\cite{ATLAS-CONF-2013-053},
ATLAS-CONF-2013-062~\cite{ATLAS-CONF-2013-062},
ATLAS-CONF-2013-065~\cite{ATLAS-CONF-2013-065},
SUSY-2013-05~\cite{SUSY-2013-05}
\item {\bf CMS:} 
SUS-11-022~\cite{SUS-11-022},
SUS-12-028~\cite{SUS-12-028},
SUS-13-002~\cite{SUS-13-002}, 
SUS-13-004~\cite{SUS-13-004},
SUS-13-008~\cite{SUS-13-008}, 
SUS-13-011~\cite{SUS-13-011}, 
SUS-13-013~\cite{SUS-13-013}
\end{itemize}

Of course the database is continuously being extended as new results become available.

\subsection{Matching Theoretical and Experimental results}
\label{sec:interface}

Once a BSM spectrum is decomposed according to the procedure described in Section~\ref{sec:decomposition}, 
all the relevant information for confronting the model with the experimental results is encapsulated in the SMS
topologies plus their mass vectors and weights. Any specific model dependent information can be dropped at this point.
As discussed in Section~\ref{sec:database}, 
it is however often the case that the experimental result constrains
a sum of topologies instead of a single one. 
Before a direct comparison with the experimental constraints, it is necessary
to combine single SMS topologies (which means adding their weights)
according to the experimental analysis' assumptions (described in the metadata of the analysis).  
Furthermore, it is always implicitly assumed that all summed topologies have a common BSM mass
vector (for the slepton pair production example this means $(m_{\tilde{e}},m_{\tilde{\chi}_{1}^{0}}) = (m_{\tilde{\mu}},m_{\tilde{\chi}_{1}^{0}})$). 
Therefore, when combining signal topologies according to the analysis constraints, 
we must ensure that they have {\em similar} mass vectors.

Since different analysis have different sensitivities to the mass vector,
we do not use the simple mass distance in GeV in order to verify if two vectors are
similar or not. Instead we consider the sensitivity of
the analysis in question to the difference between the two vectors.
In order to quantify this sensitivity we use the analysis upper limit 
for each individual mass vector. If both upper limits differ by less than a maximal amount (20\%),
we render the mass vectors as similar {\it with respect to this particular analysis}.
However, if the upper limits differ by more than 20\%,
we consider the two vectors as distinct. Moreover, in order to avoid cases
where two upper limits are coincidentally equal, but they correspond to completely different mass configurations,
we also require the mass values not differ by more than 100\%.

Once again we illustrate this procedure using the slepton pair production constraint as an example.
In the MSSM we usually have $m_{\tilde{e}_L} = m_{\tilde{\mu}_L}$ and 
$m_{\tilde{e}_R} = m_{\tilde{\mu}_R}$, but $m_{\tilde{e}_L} \neq m_{\tilde{e}_R}$.
The experimental constraint on slepton pair production requires to combine the topologies $[[[e^+]],[[e^-]]]$ and 
$[[[\mu^+]],[[\mu^-]]]$ with $(m_{\tilde{e}},m_{\tilde{\chi}_{1}^{0}}) \simeq (m_{\tilde{\mu}},m_{\tilde{\chi}_{1}^{0}})$. Since $\tilde{e}_{L,R}^+\tilde{e}_{L,R}^-$ and  $\tilde{\mu}_{L,R}^+\tilde{\mu}_{L,R}^-$
contribute to $[[[e^+]],[[e^-]]]+[[[\mu^+]],[[\mu^-]]]$, we must first group the sleptons ($\tilde{e}_{L}$, $\tilde{e}_{R}$, $\tilde{\mu}_{L}$, $\tilde{\mu}_{R}$) with similar masses before we can combine the topologies.
In order to identify the similar mass vectors we first obtain, {\it for the given analysis}, the upper limit for each vector ($(m_{\tilde{e}_L},m_{\tilde{\chi}_{1}^{0}})$, $(m_{\tilde{\mu}_R},m_{\tilde{\chi}_{1}^{0}})$,...)
and cluster together the masses with similar upper limit values. If the analysis is sensitive to the
left-handed/right-handed slepton mass splitting, the upper limits will differ significantly and the
grouped masses will correspond to ($\tilde{e}_L$,$\tilde{\mu}_L$) and ($\tilde{e}_R$,$\tilde{\mu}_R$).
On the other hand, if the mass spliting is small and the analysis is not sensitive to it,
all upper limits will be similar and all the sleptons will be grouped together.

After the topologies with similar mass vectors have been identified, 
we can combine them (add their $\sigmaXBF$) according to the experimental constraint
($\sigma([[[e^+]],[[e^-]]])+\sigma([[[\mu^+]],[[\mu^-]]])$ for the example above).
However, as mentioned in Section~\ref{sec:database}, for most constraints involving a sum of single topologies,
the experimental assumptions include conditions on each topology contributing to the sum, such
as $\sigma([[[e^+]],[[e^-]]]) \simeq \sigma([[[\mu^+]],[[\mu^-]]])$ for the slepton analysis.
These conditions need to be taken into account 
when interpreting the experimental results, since each topology
may have a different signal efficiency.
Therefore we must also verify that these conditions are satisfied, otherwise
the experimental upper limit can not be applied.
Finally, if the experimental assumptions are satisfied, the resulting theoretical predictions (\sigmaXBF) 
obtained after combining the topologies can be directly compared to the
corresponding experimental upper limit. 

In Fig.~\ref{fig:scheme} we summarize the main steps required to confront the BSM model predictions 
with the experimental constraints: 
the SMS decomposition, the combination of SMS topologies with identical or similar masses 
into the topology sums assumed by the 
analyses and finally the comparison with the experimental upper limits obtained from the 
database of LHC results described in Section~\ref{sec:database}.

\section{Validation} \label{sec:validation}

A simple and robust way to validate the {\tt SModelS} procedure described in Section~\ref{sec:details}
consists in applying it to a simplified model. In this case, the experimental assumptions
are exactly satisfied by the full (simplified) model and we should be able to
reproduce the exclusion curve given by the experimental collaboration.
For instance, to validate the ATLAS-CONF-2013-035 
$\tilde\chi^\pm_1\tilde\chi^0_2 \to WZ\tilde{\chi}_1^0\tilde{\chi}_1^0$
analysis, we assume the full model only consists of a pair of mass-degenerate
$\tilde\chi^\pm_1$ and $\tilde\chi^0_2$ (both pure Winos) and the neutralino
LSP (pure Bino), $\tilde\chi^0_1$, with ${\cal B}(\tilde\chi^\pm_1\to
W^{(*)}\tilde{\chi}_1^0) = {\cal B}(\tilde\chi^0_2\to Z^{(*)}\tilde{\chi}_1^0) = 1$.
All the other particles in the spectrum are taken to be in the multi-TeV scale and decoupled. 
Scanning over $m_{\tilde\chi^\pm_1}$ and $m_{\tilde{\chi}_1^0}$ and verifying which points are excluded
by the ATLAS-CONF-2013-035 analysis, we then expect to reproduce the
official exclusion curve obtained by the experimental
collaboration.\footnote{Since we compute the EW gauginos and
sleptons cross sections at LO, while the experimental collaborations assume NLO
cross sections, we multiply the LO cross sections by a constant K-factor (1.2)
to more accurately reproduce the official curve.}

\begin{figure}\centering
  \includegraphics[width=0.8\textwidth]{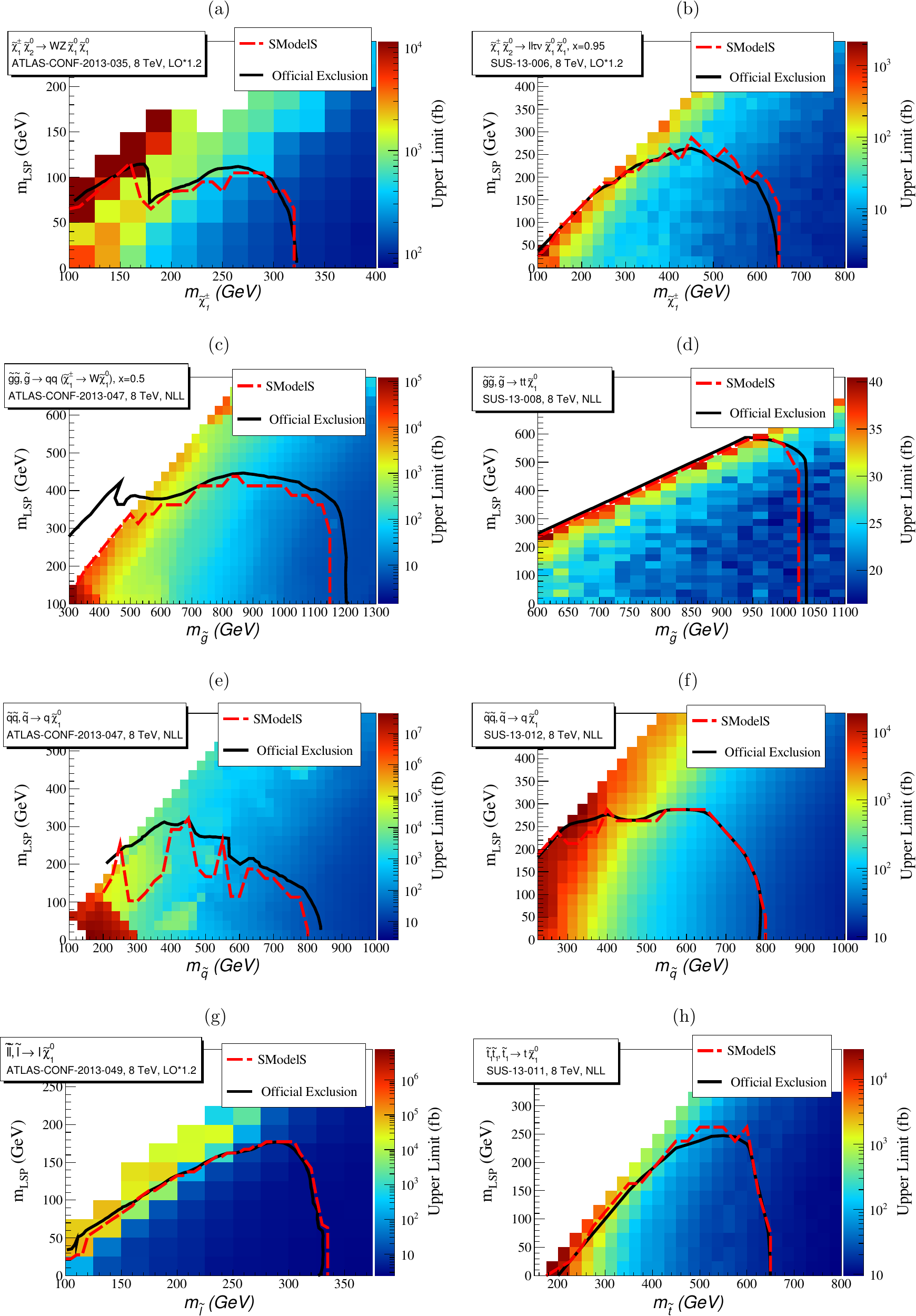}
\caption{\label{fig:RegressionTest}
Examples for regression tests for some ATLAS (on the left) and CMS (on the right) analyses. See text for details.}
\end{figure}

We performed such ``regression tests'' for several SMS results implemented
in the {\tt SModelS} database.
Some illustrative examples are shown in Fig.~\ref{fig:RegressionTest}. In particular, we show 
regression tests for ATLAS and CMS electroweak-ino searches, 
searches for gluinos with three-body decays into light quarks (ATLAS) or top quarks (CMS), 
squark pair-production with direct decay into jet+LSP (ATLAS and CMS),
slepton-pair production (ATLAS), as well as stop-pair production giving $t\bar t$+MET (CMS).
The upper limit obtained by interpolating the experimental results is shown by
the color coded background and we assumed a constant binning of 25~GeV
for all plots. The official exclusion curve obtained by the experimental
collaboration is shown in black (solid), while the one obtained through {\tt
SModelS} is shown in red (dashed). In general we find very good agreement with the official
exclusion curves once we take into account
differences in binning size and the smoothing
procedure adopted by the experimental collaboration. 
(In Fig.~\ref{fig:RegressionTest}(c), the red line and the background do not extend 
above the diagonal because the results for 
off-shell chargino decays are not yet included in the database.)
The exception is the ATLAS $\tilde q\tilde q, \tilde q\to q\tilde\chi^0_1$ analysis \cite{ATLAS-CONF-2013-047}: 
here there are regions below the official exclusion line which are actually not excluded by the 95\% CL UL 
on the cross section. This is due to large fluctuations in the upper limit values published by 
the experimental collaboration, clearly visible in the colored bins in Fig.~\ref{fig:RegressionTest}(e).

\clearpage
\section{Application to the MSSM} \label{sec:scan} 

\subsection{Parameter scans}

To demonstrate how the SMS results currently constrain supersymmetry, and to see which scenarios remain untested by the current searches (based on their SMS interpretations), we perform a scan over the MSSM with parameters defined at the weak scale. 
The parameters which we consider are the following: 
\begin{itemize}
\item the gaugino mass parameters $M_1$, $M_2$ and $M_3$, assuming the approximate GUT relation $M_1:M_2:M_3=1:2:6$;
\item the higgsino mass parameter $\mu$, the pseudoscalar mass $m_A$ and $\tan\beta=v_2/v_1$;
\item a common mass parameter $M_{\tilde q}\equiv M_{\tilde Q_{1,2}}=M_{\tilde U_{1,2}}=M_{\tilde D_{1,2}}$ for the first two generations of left- and right-handed squarks;
\item mass parameters and trilinear couplings for stops and sbottoms: $M_{\tilde Q_3}$, $M_{\tilde U_3}$, $M_{\tilde D_3}$, $A_t$, $A_b$
\item for left and right sleptons, we take common mass parameters $M_{\tilde L}$, $M_{\tilde R}$ for all three generations, supplemented by the trilinear couplings for staus, $A_\tau$;
\end{itemize}
Since we do not consider heavy Higgses in our analysis, we fix $m_A=2$~TeV. 
This leaves us with 12 free parameters. We perform two random scans over this parameter space: one focusing on gauginos 
and sleptons with all squarks assumed to be heavy ({\bf Scan-I}), and one where
sleptons are assumed to be heavy but squarks, and in particular stops and
sbottoms, can be light  ({\bf Scan-II}).  
The scan ranges and fixed parameters for the two scans are summarized in Table~\ref{tab:scan}.\footnote{In {\bf Scan-II} it is much more efficient to find points which pass all constraints, in particular the $\br(b \rightarrow s\gamma)$ constraint, if $A_t>0$ than for $A_t<0$. Since the physical observables we are interested in (\ie\ the coverage of masses and decay branching ratios) are largely insensitive to the sign of $A_t$, 
we choose to consider only positive $A_t$. Concretely we limit the scan to
$A_t>1$~TeV as for lower $|A_t|$ we obtain too low $m_h$.}

\begin{table}[h!]\centering
\caption{Scan ranges and values for fixed parameters used in this study;  dimensionful quantities are in TeV units. 
$R\equiv \max(M_{\tilde Q_3},M_{\tilde U_3})$. The parameters are scanned over randomly assuming a flat distribution. \label{tab:scan}}
\vspace*{4mm}
\begin{tabular}{l|c|c|c|c|c|c|c|c|c|c|c|c}
  & $M_2$ & $\mu$ & $\tan\beta$ & $M_{\tilde L}$ & $M_{\tilde E}$ & $M_{\tilde q}$ & $M_{\tilde Q_3}$ & $M_{\tilde U_3}$ & $M_{\tilde D_3}$ & $A_t$ & $A_b$ & $A_\tau$ \\ 
  \hline
{\bf Scan-I} & 0.1--1 & 0.1--1 & 3--60 & 0.1--1 & 0.1--1 & 5 &  2 & 2 & 2 & $\pm 6$ & $ 0 $ & $\pm 1$ \\
  \hline 
{\bf Scan-II} & 0.1--1 & 0.1--1 & 3--60 & 5 & 5& 0.1--5 &  0--2 & 0--2 & 0--2 & $[1,3R]$ & $\pm 1$ & $ 0 $ \\
\end{tabular}
\end{table}

We thus effectively have 7 free parameters in {\bf Scan-I} and 9  free parameters in {\bf Scan-II}.
We use SOFTSUSY3.3.9~\cite{Allanach:2001kg} for the computation of the masses and mixings,  
SDECAY1.3~\cite{Muhlleitner:2003vg} for the sparticle decay tables, 
SUPERISO3.3~\cite{Mahmoudi:2008tp} for flavor observables, and the {\tt masslimits} function 
of micrOMEGAs3.2\cite{Belanger:2001fz,Belanger:2004yn,Belanger:2013oya}  for LEP limits 
(we also compute dark matter observables with micrOMEGAs3.2 but do not use them here).
We require that the LSP be the $\tilde\chi^0_1$, and that all heavier sparticles decay promptly. 
Concretely we require $m_{\chipm}-m_{\chiz}>300$~MeV to veto long-lived charginos, for which the 
available SMS results do not apply. Finally, we require that points be consistent with measurements of 
$\br(b \rightarrow s\gamma)$,  \br$(B_s \rightarrow \mu^+ \mu^-)$, the muon $(g-2)$, the $Z\to{\rm invisible}$ width, 
and the mass of the SM-like Higgs boson. We also apply SUSY mass limits from LEP.   
The constraints imposed are listed in Table~\ref{tab:constraints}. 
In each scan, we collected more than 30K points which obey all these
constraints; these points are then passed through {\tt SModelS} to test their
compatibility with the SMS constraints.
{\tt SModelS} decomposes the spectrum of each point into its SMS topologies and compares it to the experimental constraints according to the procedure described in the previous sections. It returns detailed information on the 
occurring topologies, which topologies are actually tested ({\it i.e.}
SMS topologies which have been constrained by one or more experimental analyses
in the database), and how their \sigmaXBF compares to the experimental UL.
For the results presented below, we only considered the 8~TeV analyses in the database. 

\begin{table}[tb]
\caption{Constraints used to define the valid parameter space before applying the SMS limits.
\label{tab:constraints} } \vspace*{4mm}
\hspace*{-6mm}\begin{tabular}{|c|c|c|c|c|}
\hline
  \bf Observable  & \bf Experimental result   & \bf Theory uncert. & \bf Constraint imposed \\
\hline
$\br(b \rightarrow s\gamma)$ & $(3.43 \pm 0.21 \pm 0.07)\times 10^{-4}$ \cite{Amhis:2012bh} & $0.23 \times 10^{-4}$ \cite{Misiak:2006ab} & $[2.79,\,4.07] \times 10^{-4}$  \\
\hline
\br$(B_s \rightarrow \mu^+ \mu^-)$ & $(2.9\pm 0.7) \times 10^{-9}$ \cite{CMS-PAS-BPH-13-007} & 10\% & $[1.4,\,4.4] \times 10^{-9}$\\
\hline
$\Delta a_\mu$ & $(26.1 \pm 8.0)\times 10^{-10}$ \cite{Hagiwara:2011af} & $\sim 8\times 10^{-10}$ & $<5\times 10^{-9}$ \\   
\hline
$\Delta\Gamma(Z\to{\rm inv})$ & & --- & $<3$~GeV \\
\hline
$m_h$ & $125.5\pm 0.2\,^{+0.5}_{-0.6}$ GeV (ATLAS) \cite{ATLAS-CONF-2013-014} 
              & 3~GeV \cite{Allanach:2004rh,Kant:2011mq}   & $125.5\pm 3$~GeV \\
              & $125.7\pm 0.3\pm 0.3$ GeV \,(CMS) \cite{CMS-PAS-HIG-13-005} & &  \\
\hline
 sparticle masses & LEP  & --- & (micrOMEGAs~\cite{Belanger:2001fz,Belanger:2004yn})\\
\hline
\end{tabular}
\end{table}

\subsection{Results from Scan-I (EW-ino and slepton focus)}

\begin{figure}[t!]
  \includegraphics[width=\textwidth]{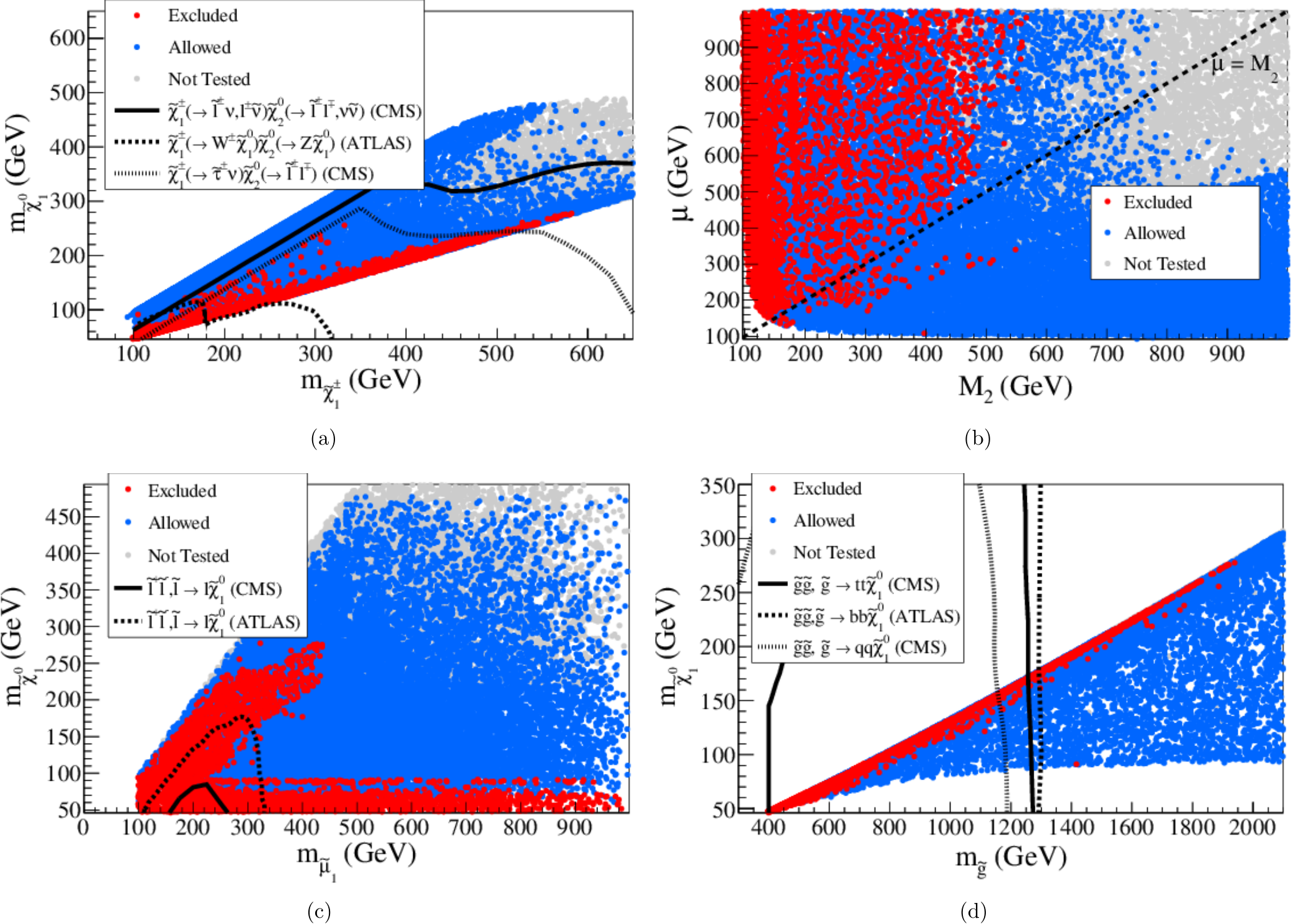} 
\caption{Excluded (red), not excluded (blue) and not tested (grey) points from {\bf Scan-I} 
in the a) top left: $m_{\tilde\chi^0_1}$ versus $m_{\tilde\chi^\pm_1}$, b) top right: $M_2$ versus $\mu$ , 
c) bottom left: $m_{\tilde \mu_1}$ versus $m_{\tilde\chi^0_1}$ and d) bottom right: $m_{\tilde g}$ versus 
$m_{\tilde\chi^0_1}$ planes. The solid, dashed and dotted black lines show the official exclusions curves 
from particular ATLAS and CMS analyses as explained in the text.
\label{fig:ResultsScan1Excall}}
\end{figure}

The purpose of {\bf Scan-I} is to test the sensitivity to EW-ino and slepton searches. Moreover, as we assume 
a GUT relation between the gaugino masses, it will also be susceptible to limits from three-body gluino decays. 
To begin with, we show in Fig.~\ref{fig:ResultsScan1Excall} scatter plots 
of the scan points in the $m_{\tilde\chi^\pm_1}$ versus $m_{\tilde\chi^0_1}$, $M_2$  versus $\mu$, 
$m_{\tilde \mu_1}$  versus $m_{\tilde\chi^0_1}$ and $m_{\tilde g}$  versus $m_{\tilde\chi^0_1}$ planes. 
The red points, which form the top layer, are excluded by at least one analysis, while the 
blue points are not excluded by any single SMS limit. 
The grey points do not have any experimental limits either because their topologies do not match 
any of the existing experimental results or because their masses fall outside the
ranges considered by the experimental searches.
These points are labeled `not tested' and mostly appear when
both the chargino and gluino masses fall outside the grids of the experimental results. 

For comparison purposes we also show official 95\% CL exclusion curves for some SMS topologies. 
In Fig.~\ref{fig:ResultsScan1Excall}(a), the $m_{\tilde\chi^0_1}$ versus $m_{\tilde\chi^\pm_1}$ plot, 
we show the exclusion curves for $\tilde{\chi}_{1}^{\pm}\tilde{\chi}_{2}^{0}$
production with democratic decays to sleptons and sneutrinos assuming
$m_{\tilde l} = m_{\tilde \nu} = (m_{\tilde{\chi}_{2}^{0}} +
m_{\tilde{\chi}_{1}^{0}})/2$ (Fig.~14 of \cite{SUS-13-006}),
with $\tau$-enriched decays and $m_{\tilde l} = (0.95 m_{\tilde{\chi}_{2}^{0}} +
0.05 m_{\tilde{\chi}_{1}^{0}})$ (Fig.~16a of \cite{SUS-13-006}), 
and with decays through on or off-shell $W,Z$ (Fig.~8b of  \cite{ATLAS-CONF-2013-035}).
The exclusion curves for slepton pair production and decay 
($\tilde l_{L,R}^+ \tilde l_{L,R}^- \to l^+l^- \tilde \chi_1^0 \tilde \chi_1^0$)
for both ATLAS and CMS (Figs. 16a from \cite{ATLAS-CONF-2013-049} and 20b from \cite{SUS-13-006})
 are shown in Fig.~\ref{fig:ResultsScan1Excall}(c), \ie\ the plot of $m_{\tilde \mu_1}$ versus $m_{\tilde\chi^0_1}$. Finally, in Fig.~\ref{fig:ResultsScan1Excall}(d), the projection onto the $m_{\tilde g}$--$m_{\tilde\chi^0_1}$ plane, we show the SMS exclusion curves for gluino topologies with 3-body decays to the LSP and light quarks (Fig.~6b of \cite{SUS-13-012}), $\bar{t}t$ (Fig.~13 of \cite{SUS-13-004}) or $\bar{b}b$ (Fig.~12a of  \cite{ATLAS-CONF-2013-061}).  

Some comments are in order regarding the differences between the areas covered by the red points and 
the naive expectations from the official exclusion curves. One issue concerns the gaugino--higgsino mixing. 
As seen in Figs.~\ref{fig:ResultsScan1Excall}(a) and \ref{fig:ResultsScan1Excall}(d), 
excluded points are concentrated along the $m_{\tilde{\chi}_{1}^{\pm}} \sim 2m_{\tilde{\chi}_{1}^{0}}$ 
(or $m_{\tilde g} \sim 7m_{\tilde{\chi}_{1}^{0}}$)  line, which corresponds
to model points with a pure wino chargino ($\tilde \chi_1^{\pm} \simeq \tilde{W}^{\pm}$).
Once $\tilde{\chi}_{2}^{0}$ and $\tilde{\chi}_{1}^{\pm}$ acquire a higgsino component,
the constraints become significantly weaker. Notice that this happens even for points with
large $m_{\tilde{\chi}_{1}^{\pm}}-m_{\tilde{\chi}_{1}^{0}}$ mass splittings, far from the pure 
higgsino region.
This is explicitly visible in Fig.~\ref{fig:ResultsScan1Excall}(b), which shows the projection onto the 
$M_2$--$\mu$ plane. As we can see, almost all excluded points lie in the wino chargino region ($\mu > M_2$).

In Fig.~\ref{fig:ResultsScan1Excall}(c) we show the same results, but now in the $\tilde \mu_1$--$\tilde\chi^0_1$ mass plane, where we see that for $m_{\tilde \chi_1^0} \lesssim 125$ GeV there are excluded points for any slepton mass, due to constraints on light charginos (decaying to $WZ$) and gluinos. Once the LSP becomes heavier, the constraints for $\tilde{\chi}_{1}^{\pm}\tilde{\chi}_{2}^{0}$ decaying through sleptons become relevant and exclude the region $m_{\tilde \mu_1}  \lesssim 2m_{\tilde \chi_1^0}$ up to $m_{\tilde \chi_1^0} \sim 250$ GeV ($m_{\tilde \chi_1^{\pm}} \sim 500$ GeV).
Finally, in Fig.~\ref{fig:ResultsScan1Excall}(d),  the gluino versus neutralino mass plane, we see once again that the points are concentrated along the bino LSP line ($m_{\tilde g} \sim 7m_{\tilde{\chi}_{1}^{0}}$).

\begin{figure}[t!]
  \includegraphics[width=\textwidth]{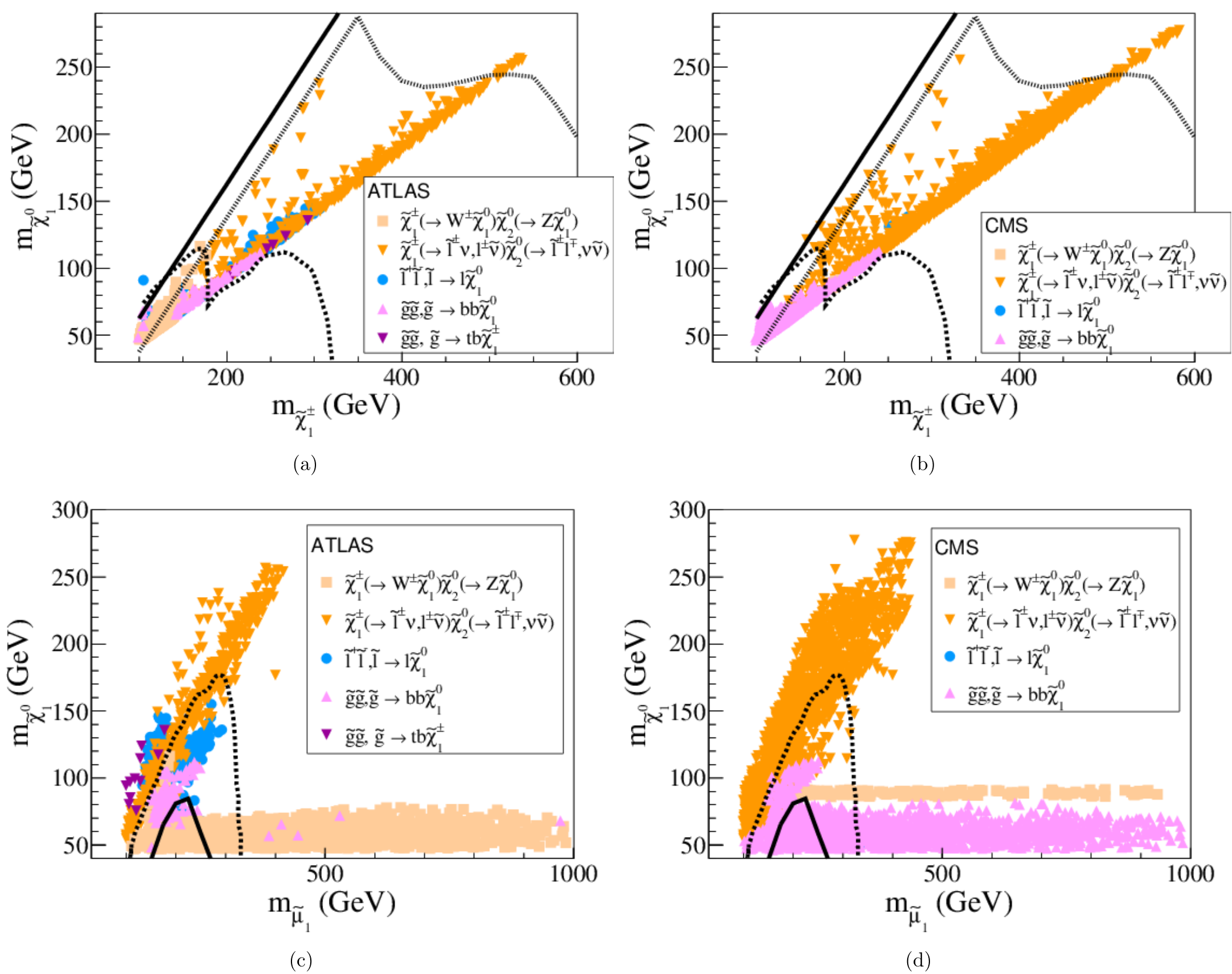} 
\caption{Breakdown of the most constraining analyses for each excluded parameter point from {\bf Scan-I}, on the left 
for ATLAS and on the right for CMS results. The top row shows the $\tilde \chi_1^\pm$ versus $\tilde \chi_1^0$, 
the bottom row shows the $\tilde \mu_1$ versus $\tilde \chi_1^0$ mass planes.
The labels correspond to the SMS topology constraining the respective point. If not shown explicitly, charginos,
neutralinos, sleptons and sneutrinos are assumed to decay as $\tilde \chi_1^\pm \to W^{\pm}\tilde \chi_1^0$, $\tilde \chi_2^0 \to Z\tilde \chi_1^0$,
$\tilde l \to l\tilde \chi_1^0$ and $\tilde \nu \to \nu\tilde \chi_1^0$, respectively.
The exclusion curves are the same as in Figs.~\ref{fig:ResultsScan1Excall}(a) and \ref{fig:ResultsScan1Excall}(c).
\label{fig:ResultsScan1} }
\end{figure}

It is also interesting in this respect to look at a more detailed breakdown 
of how different analyses constrain the parameter space.
In Fig.~\ref{fig:ResultsScan1} we show the most constraining topology for each
{\it excluded} point in the chargino--neutralino and slepton--neutralino mass planes.\footnote{By most constraining we here mean the topology that gives the largest ratio of predicted $\sigmaXBF$ over the 95\% UL $\sigmaXBF$. }
We also split the constraints by ATLAS (left panel) and CMS (right panel) results.
The first interesting point to notice is that the constraints coming from gluino decay topologies 
(mostly from $\tilde g\tilde g$ production with $\tilde g\to b\bar b\tilde\chi^0_1$) 
are stronger then those from EW production only at low chargino masses ($m_{\tilde\chi^\pm_1}
\lesssim 250$ GeV), while the high mass region is constrained by either slepton
pair production or $\tilde{\chi}_{1}^{\pm}\tilde{\chi}_{2}^{0}$ production
followed by decay through on-shell sleptons. 
Both ATLAS and CMS results show a similar behavior.\footnote{In the right-hand side (CMS) plots 
of Fig.~\ref{fig:ResultsScan1}, the pink points excluded by gluino topologies are shown as the bottom layer 
and are hence partly covered by the yellow points excluded by EW-ino topologies, but they do extend up to 
$m_{\tilde\chi^\pm_1}\lesssim 250$ GeV.} 
The main differences come from the ATLAS off-shell WZ analysis, which can extend the constraints
to a significant part of the mixed-chargino region (at low chargino masses), and from 
the ATLAS $\tilde l \tilde l$ search, which obtained stronger
constraints than the equivalent CMS analysis. 
On the other hand, the 2012 CMS EW-ino analysis \cite{SUS-12-022} 
saw an under-fluctuation in the BG, which resulted in
stronger constraints for some regions of parameter space than the ATLAS results.\footnote{In principle one should of course use only the constraints from the most sensitive analysis, 
in order not to alter the effective CL. Ideally, this should be based on the {\em expected} limits. 
Since the expected limits are however not provided by the experimental collaborations 
systematically for all results, for the time being we choose to use all constraints in a democratic manner. We hope that situation will improve as SMS results become more widely used (see, \eg\, the recent LPCC workshop \cite{SMSworkshop}). }

\begin{figure}[t!]\centering
\includegraphics[width=0.49\textwidth,clip]{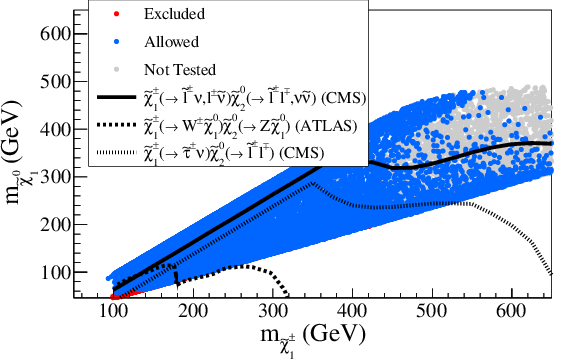}
\includegraphics[width=0.49\textwidth,clip]{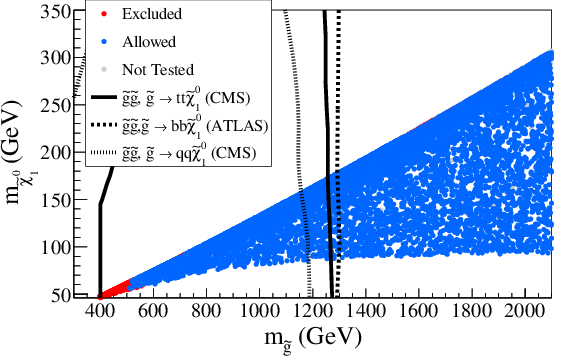}
\caption{SMS-allowed points from  {\bf Scan-I} in the chargino versus neutralino mass plane (left) and gluino versus neutralino mass plane (right). The exclusion curves are the same as in Figs.~\ref{fig:ResultsScan1Excall}(a) and \ref{fig:ResultsScan1Excall}(d).
\label{fig:ResultsScan1Allowall} }
\end{figure}

We note furthermore that points with the same ($m_{\tilde\chi^\pm_1}, m_{\tilde\chi^0_1}$) combination may 
or may not be excluded depending on whether the EW-inos are gaugino- or higgsino-like, or whether their 
decays proceed  through  $W$, $Z$, $h$ or on-shell sleptons. 
In order to verify which region of parameter space is excluded
{\it for all values of the scanned parameters}, we show in
Fig.~\ref{fig:ResultsScan1Allowall} the `allowed' points (\ie\ points that are not excluded by any of the SMS results) 
on top of the excluded ones in the $m_{\tilde{\chi}_{1}^{\pm}}$  versus $m_{\tilde{\chi}_{1}^{0}}$ 
and $(m_{\tilde g}$  versus $m_{\tilde\chi^0_1})$ planes.
We can see that only a very small fraction of points along the pure
wino-chargino line is strictly excluded by the SMS results.
The region with $m_{\tilde g} < 500$ GeV ($m_{\tilde \chi_1^0} < 70$ GeV), excluded
for any combinations of parameters, is mainly constrained by
limits on  $\tilde{\chi}_{1}^{\pm}\tilde{\chi}_{2}^{0} \to W^*Z^* \tilde \chi_1^0 \tilde \chi_1^0$,
as seen in Fig.~\ref{fig:ResultsScan1}(c). For basically any point with $m_{\tilde g} > 500$ GeV
there is at least one combination of parameters which evades all SMS constraints.
However, the official exclusion curves seem to indicate that gluino masses up to $\sim 1.2$ TeV are 
excluded by SMS constraints on gluino signal topologies.

\begin{figure}[t!]
  \includegraphics[width=\textwidth]{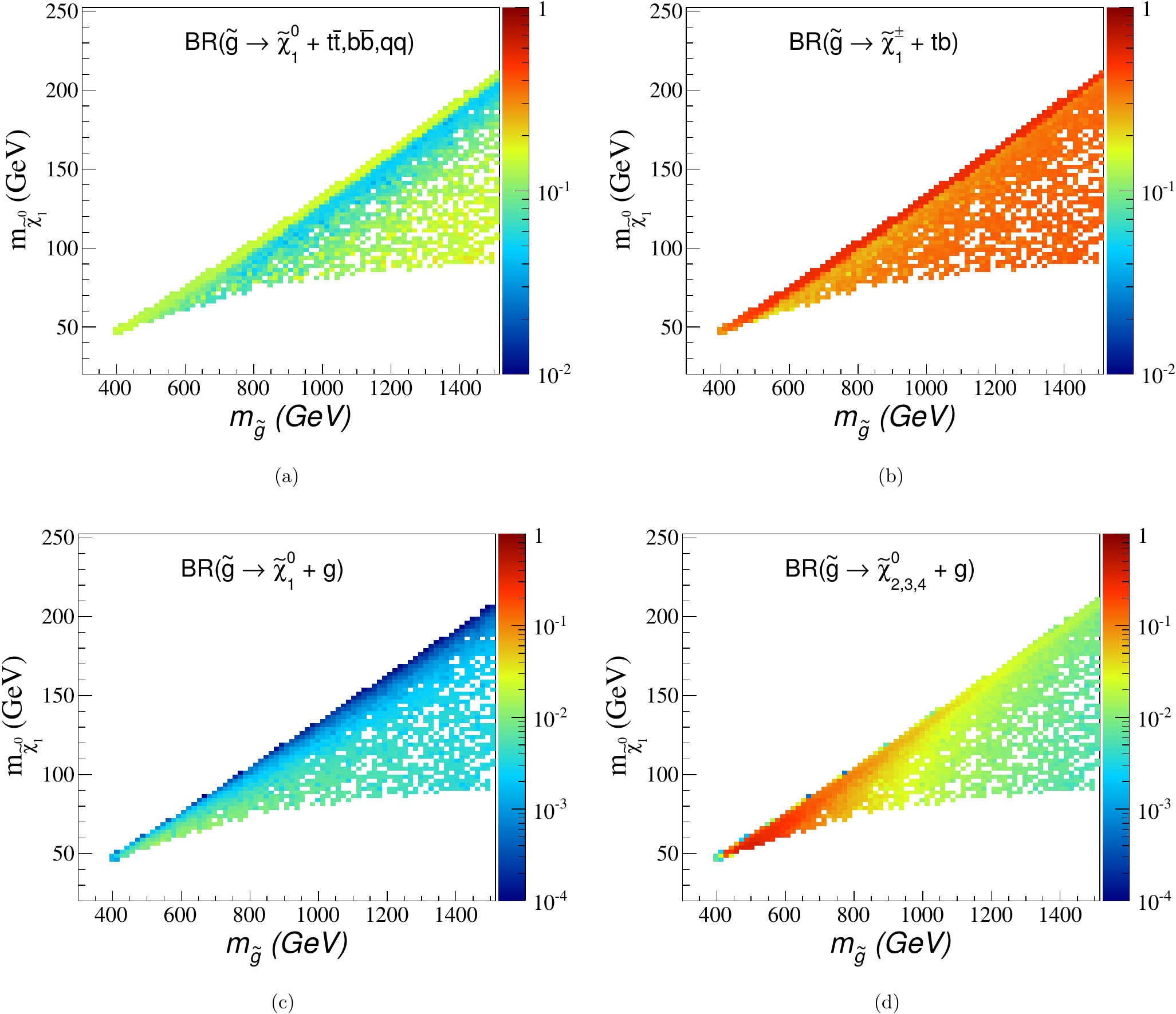} 
\caption{Maximum gluino decay branching ratios in the gluino versus neutralino mass plane for all points in {\bf Scan-I}
for the $\tilde g \to \tilde\chi_1^0 + tt,bb,qq$ (top left) and $\tilde g \to \tilde\chi_1^\pm + tb$ (top right), 
$\tilde g \to \tilde\chi_1^0 + g$ (bottom left) and $\tilde g \to \tilde\chi_{2,3,4}^0 + g$ (bottom right) decays.
\label{fig:gluinoBRs}}
\end{figure}

In order to understand how points with light gluinos evade the SMS constraints, we note that
the current SMS results for gluinos only consider direct 2-body decays to on-shell squarks or 
direct 3-body decays to the LSP plus $t\bar{t}$, $b\bar{b}$ or $q\bar q$.\footnote{A few analyses (see Refs.~\cite{SUS-13-013} and \cite{ATLAS-CONF-2013-007}, for example) consider
decays to quarks and charginos, but only for $m_{\tilde\chi^\pm_1} \simeq m_{\tilde\chi^0_1}$ or 
for decay to light quarks. Both cases are not relevant for most of the parameter space in {\bf Scan-I}.}
Since for light gluinos ($m_{\tilde g} < m_{\tilde t} \simeq 2$ TeV) the decay to on-shell squarks is kinematically forbidden, only the SMS constraints on direct decays to the LSP are relevant for {\bf Scan-I}.
Hence the excluded region will be drastically reduced
whenever the $\tilde g \to \tilde\chi_1^0 + X$ branching ratio (BR) is suppressed.
In Fig.~\ref{fig:gluinoBRs}, top left plot, 
we show the maximum value of BR$(\tilde g \to \tilde\chi_1^0 + \bar{t}t,\bar{b}b,qq)$ 
for each parameter point in the gluino versus neutralino mass plane.
As we can see, all points have BR$(\tilde g \to \tilde\chi_1^0 + \bar{t}t,\bar{b}b,qq) < 20\%$
and the region with the highest values and light gluinos ($m_{\tilde g} < 1$ TeV) 
coincide with the excluded region in Fig.~\ref{fig:ResultsScan1Excall}(d).
Since the exclusion curves always assume $100\%$ BRs, the smaller BRs lead
to an exclusion region significantly smaller than the ones projected by the curves.
We also show in Fig.~\ref{fig:gluinoBRs} the maximum of BR$(\tilde g \to \tilde\chi_1^\pm + tb)$, 
which clearly dominates over the other 3-body decays. 
However, this decay leads to gluino signal topologies not constrained by the current SMS results.
Last but not least, Fig.~\ref{fig:gluinoBRs} also shows that the loop decay into a neutralino plus a gluon jet 
can be important. 
The direct decay into the LSP, $\tilde g\to \tilde\chi_1^0+g$, is constrained by the 2 jets + MET searches. 
However, its BR is always small, see the bottom left plot. The $\tilde g\to \tilde\chi_{2,3,4}^0+g$ decays are 
much more relevant, see the bottom right plot, but as the $\tilde\chi_{2,3,4}^0$ decays will lead to additional jets 
or leptons, these are again not constrained by the current SMS results.

\begin{figure}[t!]\centering
  \includegraphics[width=0.56\textwidth,clip]{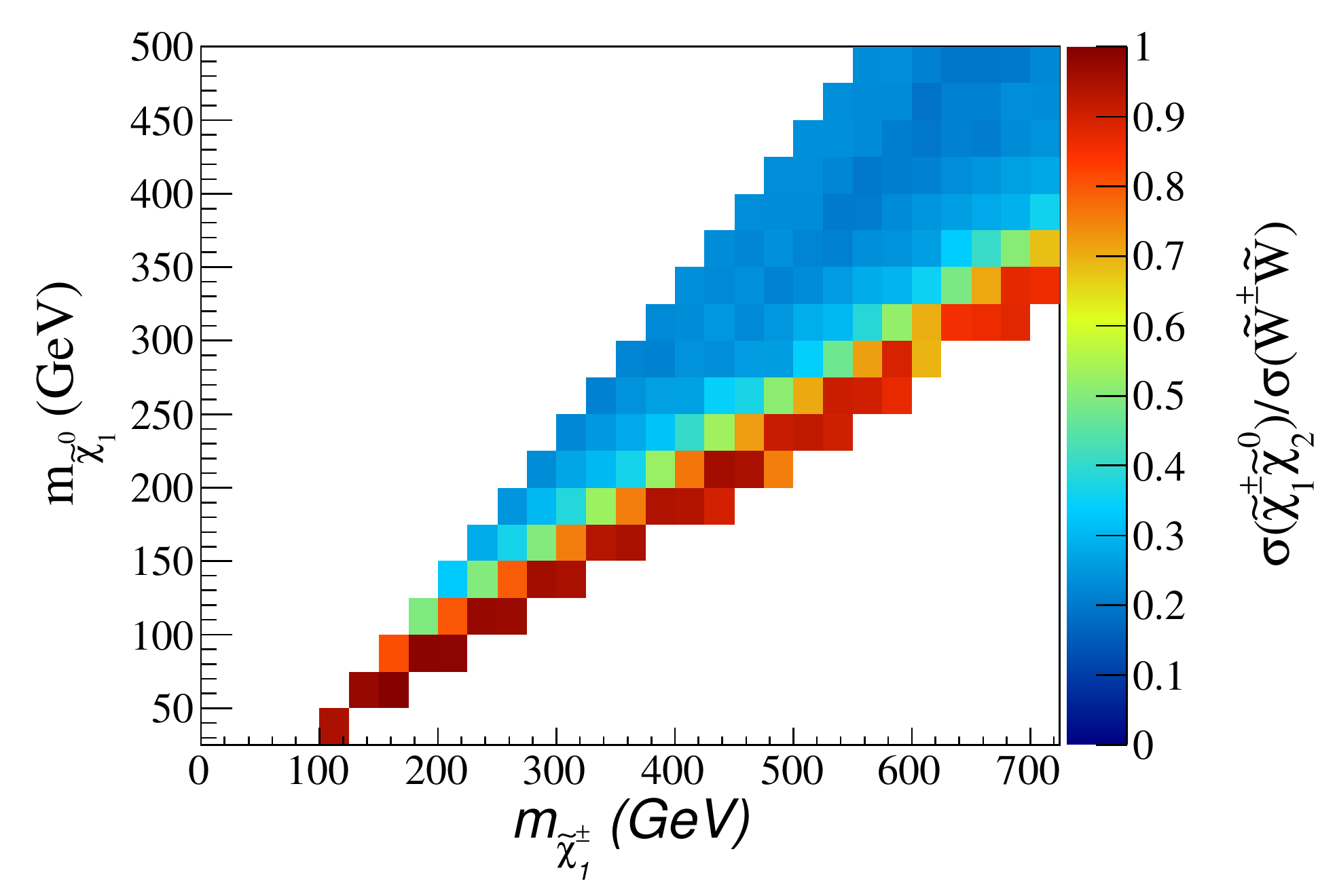}
\caption{Ratio of $\sigma(\tilde \chi_1^\pm \tilde \chi_2^0)$ and the
pure wino production cross section value ($\sigma(\tilde W^\pm \tilde W)$) 
in the chargino-neutralino mass plane for {\bf Scan-I}.
\label{fig:C1N2xsec}}
\end{figure}

Although the constraints from gluino topologies are suppressed due to the small BRs,
the same is not expected for the $\tilde \chi_1^\pm \tilde \chi_2^0$ signal topologies, since 
$\tilde \chi_1^\pm$, $\tilde \chi_2^0$ almost always decay
to on/off-shell $W$'s, $Z$'s, $h$'s or sleptons/sneutrinos and the majority of these topologies
are constrained by at least one experimental analysis.
However, as shown in Fig.~\ref{fig:ResultsScan1Excall}(b), most of the excluded points
are in the wino chargino region ($\mu > M_2$), while the mixed higgsino--wino and pure higgsino chargino
regions remains largely unchallenged. This is mainly due to the smaller higgsino production cross sections,
which suppress  $\tilde \chi_1^\pm \tilde \chi_2^0$ production once $\mu \lesssim M_2$.
This is explicitly shown in Fig.~\ref{fig:C1N2xsec} where we plot
the ratio of the $\tilde \chi_1^\pm \tilde \chi_2^0$ production cross section
and the pure wino cross section in the chargino versus neutralino mass plane.
As we can see, once we deviate from the pure wino scenario,
$\sigma(\tilde \chi_1^\pm \tilde \chi_2^0)$ decreases,
reaching values as low as 20\% of the pure wino cross section.
Although the official exclusion curves suggest that a large fraction of the mixed
higgsino-wino region is excluded, they have been obtained under the assumption
of pure wino production cross sections and do not take into account the
cross-section suppression due to the higgsino mixing.
Furthermore, points with light right-handed sleptons or light staus can also easily evade the SMS constraints.
In these cases, the
decay of neutralinos and charginos to $\tau$'s is enhanced, while the decay to $e$'s and $\mu$'s
is suppressed, resulting in a reduction of the signal efficiency for most
experimental searches. Also, once $m_{\tilde{\chi}_{2}^{0}}-m_{\tilde{\chi}_{1}^{0}} \gtrsim 125$ GeV,
the neutralino decay to $h+{\rm LSP}$ becomes the dominant decay mode. Although there are experimental analyses
which look for $\tilde{\chi}_{1}^{\pm}\tilde{\chi}_{2}^{0} \to Wh+\tilde{\chi}_{1}^{0}\tilde{\chi}_{1}^{0}$,
these are currently too weak to constrain models with gaugino mass unification.

Finally, the constraints from direct slepton production can also be suppressed
when $m_{\tilde l_R} \ll m_{\tilde l_L}$, since right-handed sleptons have smaller production
cross sections. Hence we conclude that
the parameter space excluded by the current SMS results can be
drastically reduced when compared to the naive expectations from the official exclusion curves.
For the slepton and EW-ino signal topologies 
this reduction is mostly due to the suppression of production cross sections and/or small
sensitivity to some signal topologies (such as neutralino decays to Higgs).
On the other hand, the constraints on gluino production can be potentially enhanced
if new analyses include SMS results for the most relevant gluino topology:
$\tilde g \to \tilde\chi_1^\pm + tb \to \tilde\chi_1^0 + W^{\pm} + tb$.

Last but not least note that we do not combine limits from distinct SMS topologies, which
corresponds to a conservative estimate of the excluded region. 
In particular our results tend to underestimate the exclusion obtained from gluino production: 
as explained above 
the light gluino scenarios marked as ``allowed'' in our plots typically show a rather complicated mix 
of gluino three-body decays into $\tilde\chi^0_{1,...,4}+q\bar q$ and  $\tilde\chi^\pm_{1,2}+q\bar q'$,  
and of loop-induces decays into $\tilde\chi^0_{1,...4}+g$, such that the SMS limits may be evaded 
despite a production cross section of the order of 1~pb. 
At least part of these points are effectively excluded by the generic multi-jet and/or 
multi-jet plus leptons searches.\footnote{We thank Lukas Vanelderen and Matthias Schroeder for explicit checks 
against the CMS RA2 results.} 
A way to improve this situation may be the use of efficiency maps, as discussed at the recent LPCC 
workshop~\cite{SMSworkshop}. 

\FloatBarrier

\subsection{Results from Scan-II (gluino and squark focus)}

\begin{figure}[t!]
  \includegraphics[width=\textwidth]{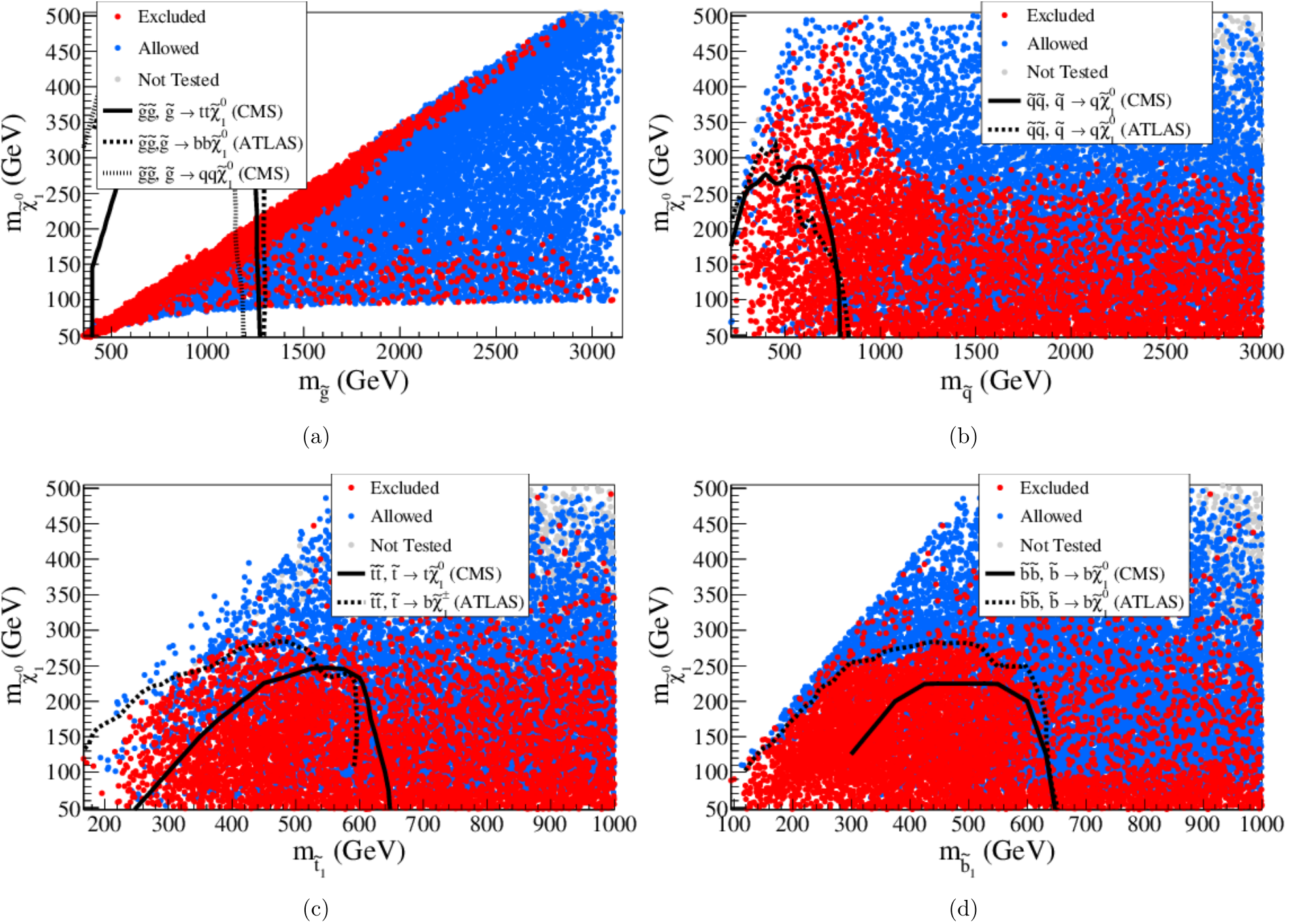} 
\caption{Excluded (red), not excluded (blue) and not tested (grey) points from {\bf Scan-II} in the 
a) top left: $m_{\tilde\chi^0_1}$ versus $m_{\tilde g}$, 
b) top right: $m_{\tilde\chi^0_1}$ versus $m_{\tilde q}$, 
c) bottom left: $m_{\tilde\chi^0_1}$ versus $m_{\tilde t_1}$, and 
d) bottom right: $m_{\tilde\chi^0_1}$ versus $m_{\tilde b_1}$ 
planes. The solid and dashed black lines show the official exclusions curves from 
particular ATLAS and CMS analyses as explained in the text.\label{fig:ResultsScan2Excall} }
\end{figure}

Let us now turn to the case that  squarks of the 1st/2nd and also of the 3rd generation are allowed to be light. 
Here, we assume decoupled sleptons. The EW-ino spectrum still is constrained by the assumption of 
gaugino-mass unification and $\mu$ is allowed to vary within 0.1--1~TeV. 
The full range of parameters is listed in Table~\ref{tab:scan}.
In Fig.~\ref{fig:ResultsScan2Excall} we show again the excluded points (in red) on top of the SMS-allowed 
points (in blue), but now in the $m_{\tilde\chi^0_1}$ versus $m_{\tilde g}$, 
$m_{\tilde\chi^0_1}$ versus $m_{\tilde q}$, $m_{\tilde\chi^0_1}$ versus $m_{\tilde t_1}$, and 
$m_{\tilde\chi^0_1}$ versus $m_{\tilde b_1}$ planes. 
Since the first two generations of squarks are degenerate,  the average squark mass is given by 
$m_{\tilde q}\equiv (m_{\tilde u_L}+m_{\tilde u_R}+m_{\tilde d_L}+m_{\tilde d_R})/4$. 

For comparison we show moreover the following official exclusion curves: 
for gluino ($\tilde g\tilde g$) topologies, we show the CMS exclusion curves for 
$\tilde g \to qq+\tilde \chi_1^0$ (Fig.~6b of \cite{SUS-13-012}) and 
$\tilde g \to \bar{t}t+\tilde \chi_1^0$ (Fig.~13 of \cite{SUS-13-004}) as well as the 
ATLAS exclusion for $\tilde g \to \bar{b}b+\tilde \chi_1^0$ (Fig.~12a of \cite{ATLAS-CONF-2013-061}); 
regarding squark production, we show the exclusion curves for 
$\tilde q \tilde q \to qq + \tilde \chi_1^0 \tilde \chi_1^0$ from both ATLAS and CMS, Figs.~6a of \cite{SUS-13-002} and 19c of \cite{ATLAS-CONF-2013-047};  for the 3rd generation, we show the exclusion curves for 
$\tilde t \bar{\tilde t}$ with $\tilde t \to t + \tilde \chi_1^0$ from CMS (Fig.~10a of \cite{SUS-13-011})
and $\tilde t \to b + \tilde \chi_1^+ \to b + W^+ + \tilde \chi_1^0$ from ATLAS  (Fig.~6 of \cite{SUSY-2013-05}) 
and the curves for $\tilde b \bar{\tilde b} \to bb + \tilde \chi_1^0 \tilde \chi_1^0$ from both ATLAS and CMS (Fig.~4 of \cite{SUSY-2013-05} and Fig.~10c of \cite{SUS-12-028}).

From Fig.~\ref{fig:ResultsScan2Excall}(a) we once again notice that the exclusion region is mostly concentrated in 
the bino LSP region ($\mu > M_1$), despite the potential presence of points with light squarks. 
Nonetheless the excluded region in the $m_{\tilde \chi_1^0}$ versus $m_{\tilde g}$ plane
is considerably larger than in {\bf Scan-I} and it extends well beyond the gluino 3-body decays exclusion curves.
This is expected since the presence of light squarks allows points to be excluded even for very heavy gluino masses.
This is explicitly visible in Fig.~\ref{fig:ResultsScan2Excall}(b), where we see that excluded points with heavy gluinos 
($m_{\tilde g} > 1.8$ TeV or $m_{\tilde \chi_1^0} > 250$ GeV) correspond to light squark masses, $m_{\tilde q} < 1$ TeV.
It is also interesting to note that the excluded points in the squark versus neutralino mass plane extend well beyond the
exclusion curves for squark production and direct decay to the LSP, even for heavy gluino masses. 
The reason is that even a 2--3 TeV gluino is not yet decoupled but gives a t-channel contribution to squark-pair production. 
Finally, in Figs.~\ref{fig:ResultsScan2Excall}(c) and \ref{fig:ResultsScan2Excall}(d), 
the excluded points are projected onto the LSP--stop and 
LSP--sbottom mass planes. While the density of excluded points seems to depend weakly on the stop mass, 
it is concentrated in the light sbottom region, agreeing well with the expectation of the  exclusion curves, except 
for light LSP masses, where constraints from gluino and EW-ino topologies become relevant.

\begin{figure}[t!]
  \includegraphics[width=0.96\textwidth]{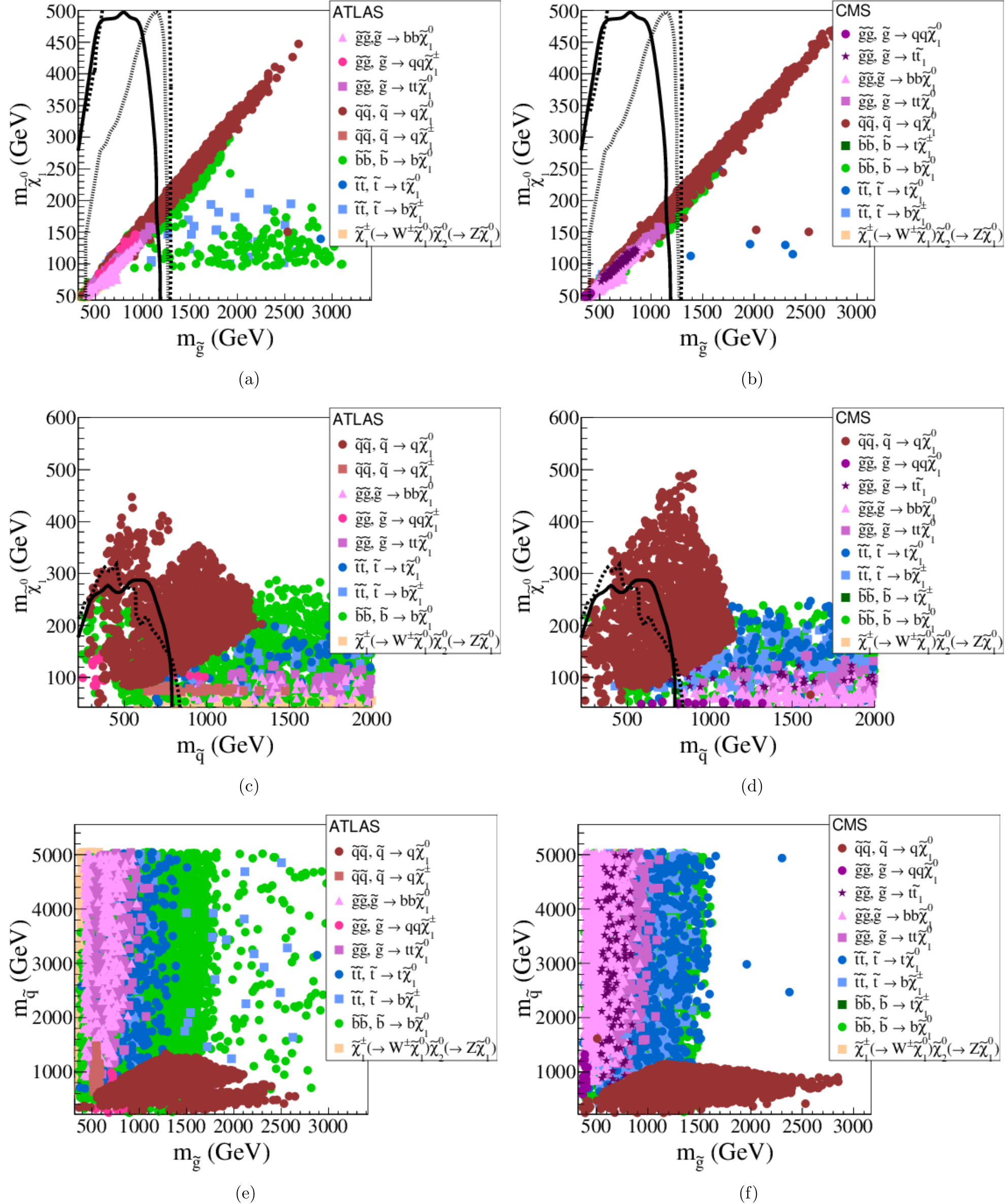} 
\caption{Breakdown of the most constraining analyses for each excluded parameter point from {\bf Scan-II}, on the left 
for ATLAS and on the right for CMS results. Shown are, from top to bottom, the LSP versus gluino, 
LSP versus squark, and squark versus gluino mass planes. 
The labels correspond to the SMS topology constraining the respective point. If not shown explicitly, charginos,
neutralinos and stops are assumed to decay as $\tilde \chi_1^\pm \to W^{\pm}\tilde \chi_1^0$, $\tilde \chi_2^0 \to Z\tilde \chi_1^0$ and
$\tilde t \to t\tilde \chi_1^0$, respectively.
The exclusion curves are the same as in the corresponding plots in Fig.~\ref{fig:ResultsScan2Excall}. 
\label{fig:ResultsBreakdown1-Scan2} }
\end{figure}

\begin{figure}[ht!]
  \includegraphics[width=\textwidth]{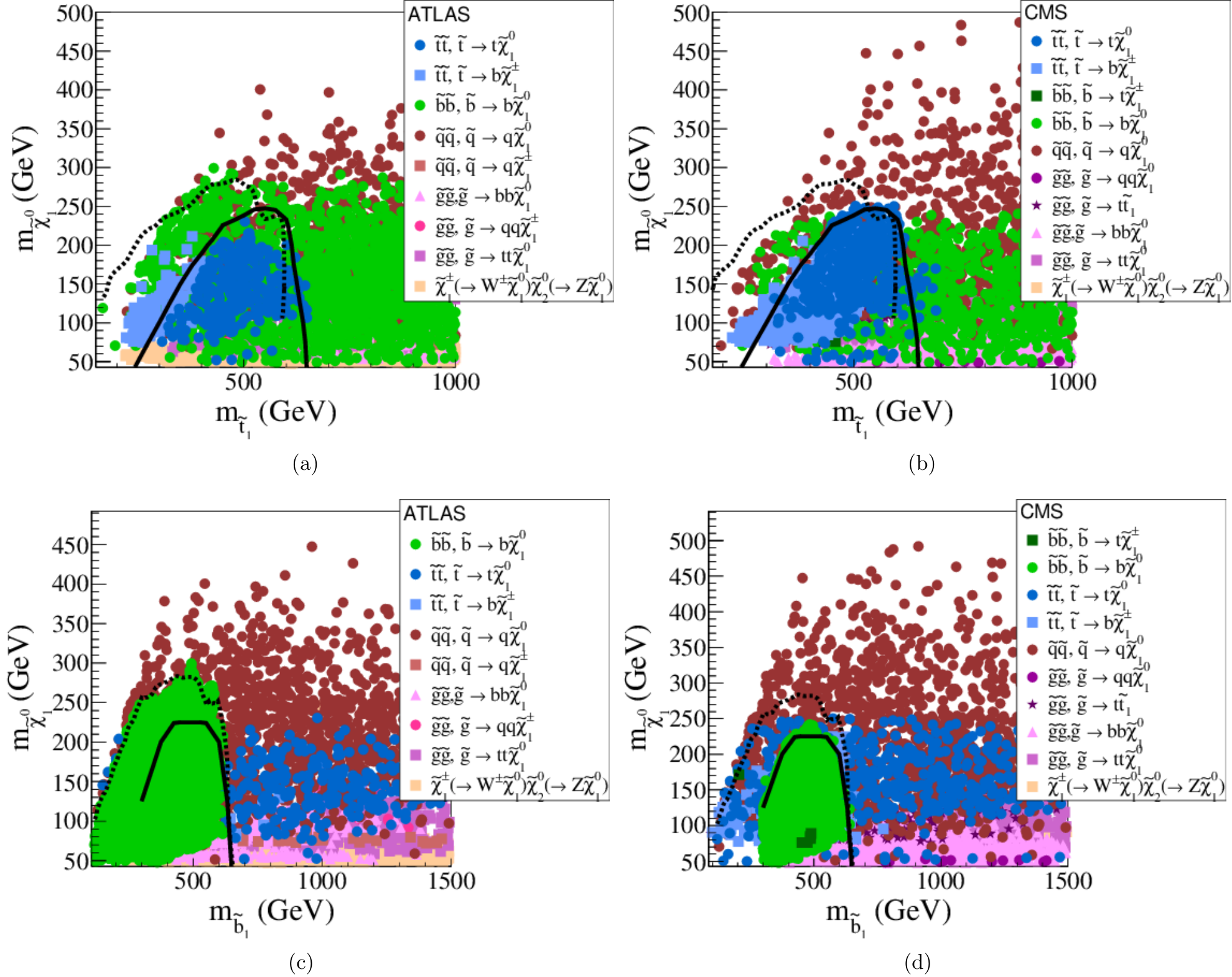} 
\caption{Breakdown of the most constraining analyses for each excluded parameter point from {\bf Scan-II} as in 
Fig.~\ref{fig:ResultsBreakdown1-Scan2} but in the $\tilde \chi_1^0$ versus $\tilde t_1$  (top row)  and $\tilde \chi_1^0$ versus $\tilde b_1$ (bottom row) mass planes.
The labels correspond to the SMS topology constraining the respective point. If not shown explicitly, charginos,
neutralinos and stops are assumed to decay as $\tilde \chi_1^\pm \to W^{\pm}\tilde \chi_1^0$, $\tilde \chi_2^0 \to Z\tilde \chi_1^0$ and
$\tilde t \to t\tilde \chi_1^0$, respectively.
The exclusion curves are the same as in the corresponding plots in Fig.~\ref{fig:ResultsScan2Excall}. 
\label{fig:ResultsBreakdown2-Scan2} }
\end{figure}

In order to better understand the excluded regions shown in Fig.~\ref{fig:ResultsScan2Excall},
we present in Figs.~\ref{fig:ResultsBreakdown1-Scan2} and \ref{fig:ResultsBreakdown2-Scan2} 
the breakdown into the most constraining topologies for each excluded point.  
We again show the constraints from ATLAS (left panels) and CMS (right panels) analyses separately. 
As already noted above, in the $m_{\tilde \chi_1^0}$ versus $m_{\tilde g}$ plane, the excluded points are concentrated 
in the bino LSP region ($m_{\tilde \chi_1^0} \sim 7 m_{\tilde g}$), see the top row of plots in 
Fig.~\ref{fig:ResultsBreakdown1-Scan2}.
The few points in the $m_{\tilde \chi_1^0} < 7 m_{\tilde g}$
region are excluded by constraints on sbottom and stop topologies, which are independent of gluino masses.
We also point out that the constraints from gluino topologies are only relevant for $m_{\tilde g} \lesssim 1.2$~TeV, while
all the points with heavier gluinos are excluded by squark topologies, as expected.
Furthermore, we see that all the points excluded by
the $\tilde q \tilde q \to qq + \tilde \chi_1^0 \tilde \chi_1^0$ topology falls along the bino LSP region.
This is expected since squarks only couple to higgsinos through their small Yukawa couplings.
Hence once the LSP acquires a sizeable higgsino component, the $\tilde q \to q + \tilde \chi_1^0$ decay 
becomes suppressed (benefitting decays into heavier neutralinos or the wino-like chargino), 
which rapidly weakens the constraints. 

In Figs.~\ref{fig:ResultsBreakdown1-Scan2}(c) and \ref{fig:ResultsBreakdown1-Scan2}(d) we show the same points, 
but in the squark versus neutralino mass plane.
The points excluded by squarks topologies ($\tilde q \tilde q \to qq \tilde \chi_1^0 \tilde \chi_1^0$) extend up
to $m_{\tilde q} \sim 1.4$ TeV, well beyond the naive expectations from the exclusion curves.
As mentioned, this is due to the t-channel gluino contribution to the squark production cross section,
which enhances $\sigma(\tilde q \tilde q)$ with respect to the fully decoupled gluino case.
Since the exclusion curves assume decoupled gluinos, the squark cross sections are significantly reduced,
resulting in a smaller reach. 
As the gluino mass increases (for a fixed squark mass), the squark cross section slowly decreases as
well as the corresponding reach in squark mass. This is explicitly seen in the plots of 
squark versus gluino mass  in Fig.~\ref{fig:ResultsBreakdown1-Scan2}. 

The same enhancement is not present for the 3rd generation squarks since the t-channel gluino contribution
is strongly suppressed by the negligible bottom and top PDFs. This is also shown by the plots in 
Fig.~\ref{fig:ResultsBreakdown1-Scan2}, where we see that the reach from sbottom and stop topologies
are independent of the gluino mass up to $m_{\tilde g} \approx 1.5-1.7$ TeV. 
For even higher gluino masses, points with a bino LSP have $m_{\tilde \chi_1^0} > 250$ GeV and are beyond the
reach of the sbottom/stop SMS constraints, as shown by the exclusion curves in 
Figs.~\ref{fig:ResultsScan2Excall}(c) and \ref{fig:ResultsScan2Excall}(d).
The few points above $m_{\tilde g} \sim 1.5-1.7$ TeV excluded by the sbottom and stop signal topologies
have a light higgsino LSP.

Finally, in Fig.~\ref{fig:ResultsBreakdown2-Scan2} we show the analyses breakdown in the stop and sbottom 
versus neutralino mass planes. We see that most of the points excluded by the stop or sbottom signal topologies 
agree well with the expectations from the exclusion curves.
One exception are the few points at $m_{\tilde t} \gtrsim 650$ GeV and $m_{\tilde\chi^0_1}\approx 150-180$~GeV 
in the top-right plot in Fig.~\ref{fig:ResultsBreakdown2-Scan2}.
These are excluded by the constraints from the CMS razor analysis~\cite{SUS-13-004}, 
which in fact has a higher reach than the exclusion curve at low LSP masses 
($m_{\tilde \chi_1^0} \lesssim 180$ GeV).
In the $m_{\tilde\chi^0_1}$ versus $m_{\tilde b_1}$ plots in 
Fig.~\ref{fig:ResultsBreakdown2-Scan2} (bottom row) the points excluded by the sbottom topologies are slightly
above the exclusion curves due to uncertainties in the production cross section, but they agree within $1\sigma$.

\FloatBarrier

\begin{figure}[t!]\centering
\includegraphics[width=0.48\textwidth]{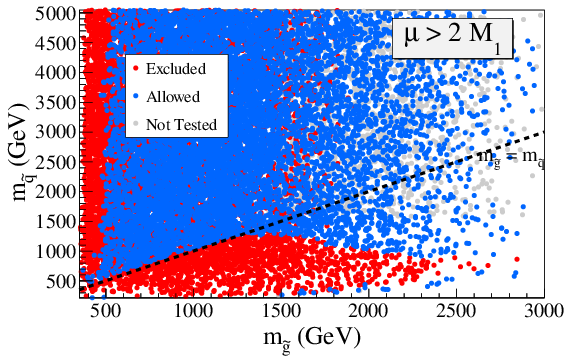} 
\includegraphics[width=0.48\textwidth]{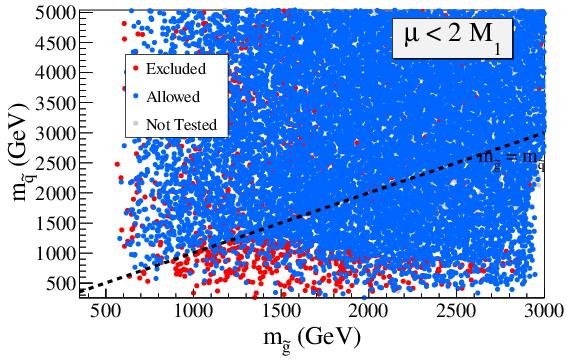}\\
\caption{SMS-allowed points from  {\bf Scan-II} in the squark versus gluino mass plane for the bino LSP case 
($\mu > 2 M_1$) and the mixed higgsino--bino scenario ($\mu < 2 M_1$). The dashed lines correspond to 
$m_{\tilde g} = m_{\tilde q}$.}
\label{fig:ResultsScan2Allowall1}
\end{figure}

We now turn our attention to the points which have large cross sections (light squark and/or gluino masses) 
but none the less evade all current SMS constraints. In Fig.~\ref{fig:ResultsScan2Allowall1} we
show the allowed points on top of the excluded ones in the squark versus gluino mass plane.
It is instructive to consider separately the bino and mixed LSP scenarios, so in the left panel we plot the
points with mostly bino $\tilde\chi^0_1$ and wino $\tilde\chi^\pm_1$ ($\mu > 2 M_1$), 
while in the right panel we show points with a mixed higgsino--bino LSP ($\mu < 2 M_1$). 
In the $\mu > 2 M_1$ case  we see that gluino masses 
below $500$ GeV are excluded for almost any choice of the parameters, 
in agreement with the results for {\bf Scan-I}.
As already mentioned, because of the relation between $M_1$, $M_2$, $M_3$, 
low gluino masses are also constrained  
by limits on $\tilde \chi_1^\pm \tilde \chi_2^0 \to WZ + \tilde \chi_1^0\tilde \chi_1^0$. 
However for any higher values of $m_{\tilde g}$ there are several allowed points as long 
as $m_{\tilde g} < m_{\tilde q}$
and/or $\mu < 2 M_1$. The former case was already discussed in the previous Section, since it is equivalent to the parameter
space of {\bf Scan-I}. As mentioned, if gluino decays to squarks are kinematically forbidden, 
the $\tilde g \to tb + \tilde \chi_1^\pm$ decay often dominates, which is not constrained by the current SMS results.
Furthermore, if squarks are heavier than gluinos, they predominatly decay to $\tilde g + q$ and these signal
topologies are also not constrained by the current SMS results. 
Therefore points with $m_{\tilde g} < m_{\tilde q}$ and $m_{\tilde g} < m_{\tilde t,\tilde b}$ can always evade the SMS constraints.
On the other hand, if $m_{\tilde g} > m_{\tilde q}$,
squarks decay to EW-inos and the constraints on squark production can be applied.
From the last plot in Fig.~\ref{fig:ResultsScan2Allowall1} we see that points with light squarks ($m_{\tilde q} \lesssim 1.3$ TeV)
and $m_{\tilde g} > m_{\tilde q}$ are always excluded in the bino LSP scenario ($\mu > 2M_1$).
Once the LSP acquires a sizeable higgsino component, the $\tilde q \to q + \tilde \chi_1^0$ decay becomes suppressed and the
constraints are much weaker, resulting in several allowed points in the light squark/gluino region, 
shown in the right plot in Fig.~\ref{fig:ResultsScan2Allowall1}.

\clearpage

\section{Conclusions}\label{sec:conclusions}

We presented a new tool, {\tt SModelS}, to decompose the LHC signatures expected 
from BSM spectra presenting a $\mathbb{Z}_2$ symmetry into SMS topologies 
and test the predicted cross sections (\sigmaXBF) for each topology against the existing 
95\% CL upper limits from ATLAS and CMS. The program consists of three parts, the 
decomposition procedure, which can be Monte-Carlo based or SLHA based, 
the database of ATLAS and CMS SMS results, and the interface to match model predictions 
onto the experimental results.  
Our concrete implementation currently focusses on SUSY searches with missing energy, 
for which a large variety of SMS results from ATLAS and CMS are available.  
{\tt SModelS} can be used ``out of the box'' for the MSSM and its extensions, like the next-to-MSSM. 
The approach is however perfectly general and easily extendible to any BSM model to 
which the experimental SMS results apply. 
(Of course, the SMS assumption is subject to 
several caveats, as explained at the beginning of Section~2 and in footnotes~3 and 4.) 

As a proof of principle, we applied our procedure in two scans of the weak-scale 
MSSM and discussed which parameter regions are excluded by the SMS limits 
and which scenarios remain untested. 
As we showed, {\tt SModelS} can be useful for several purposes. First of all it is a 
convenient tool for relatively fast surveys of complex parameter spaces in order 
to devise the regions that are definitely excluded by data. Second, it can be used to 
find relevant topologies for which no SMS results exists---this may be helpful for 
the experimental collaborations for designing new SMS interpretations.  
An example that we already identified is ``mixed'' topologies with $tb$ final states, 
which can arise \eg\ from $\tilde g \to tb\tilde\chi_1^\pm$ 
or from $\tilde t_1$ pair production with one stop decaying into $t\tilde\chi_1^0$ and 
the other one into $b\tilde\chi_1^\pm$.
Third, one may use the decomposition procedure of {\tt SModelS} independent of 
the results database in order to identify the most important signal topologies in 
regions of parameter space. 
Finally, in the case of a positive BSM signal, the {\tt SModelS} framework may be 
helpful for characterizing which new physics scenarios may explain the observations. 

While {\tt SModelS} is already a powerful and useful tool for phenomenological studies, 
there is of course still much room for improvement. For instance, as we have shown, scenarios 
with complicated decay patterns---as typical for \eg\ light gluinos and heavy squarks---are not 
well constrained with the current framework, which is based on testing the \sigmaXBF upper 
limits topology-by-topology without the possibility of combining results. 
Here the use of 
efficiency maps, once available in a systematic fashion, would allow for significant improvements. 
This is clearly a development which we will follow. 
We also foresee to extend  {\tt SModelS} to multiple branches, including sub-branches 
originating from the main cascade decay chains. This is relevant, \eg, for heavy Higgses 
(or other non-SM ``R-even'' particles as present in models with extra dimensions) appearing in the decay chains. 
We also conceive the inclusion of resonant production of new particles, as well as violation of the 
$\mathbb{Z}_2$ symmetry; in the context of SUSY this means extension to R-parity violation.  
Regarding the statistical treatment, for identifying the most sensitive analysis (in order not to alter the effective CL), 
it would be of great help if the experimental collaborations systematically provided also 
the {\em expected} \sigmaXBF upper limits in addition to the observed ones.

Last but not least note that the mapping from the full model signal to a sum over simplified model 
signal topologies is clearly not an exact procedure. It assumes that, for most experimental searches, 
the BSM model can be approximated by a sum over effective simplified models.  
The validity of this approximation, in particular {\em (i)} regarding 
the question how the type of production channel  
or the nature of the off-shell states mediating the decays affects the signal efficiencies, 
and  {\em (ii)}  regarding the question of using SMS limits derived in the SUSY context 
for non-SUSY models, will be a subject of future work. 

The publication of the {\tt SModelS} code with a dedicated online manual is in preparation~\cite{smodels}.

\FloatBarrier

\subsection*{Note added}

Shortly before submission of this paper another  program package, {\tt CheckMATE}~\cite{Drees:2013wra}, 
became available for confronting BSM scenarios with LHC Data.  {\tt CheckMATE} determines 
whether a model is excluded or not at 95\% C.L. by comparing to many recent experimental analyses, 
based on fast simulation. It thus represents an interesting alternative to the SMS approach 
followed by {\tt SModelS}. 

Moreover, while the {\tt SModelS} paper was being refereed, another program based on SMS results,  
{\tt Fastlim}~\cite{Papucci:2014rja}, was published. {\tt Fastlim} employs efficiency maps and 
currently takes into account 11 ATLAS analyses, which are mainly focused on stop and sbottom searches. 
Restricting the {\tt SModelS} database to the 11 ATLAS analyses implemented in {\tt Fastlim}, we have verified 
that for {\bf Scan-II} about $70\%$ of the points excluded by {\tt SModelS} are also excluded by 
{\tt Fastlim} and vice-versa. 
Interestingly, from all the points excluded by {\tt SModelS}, $\approx 30\%$ are not excluded by {\tt Fastlim}; 
a part of these points features a light LSP with mass of 45--65~GeV --- this region does not seem to be covered by {\tt Fastlim}. The reverse is also true: from all the points excluded by {\tt Fastlim}, $\approx30\%$ are not excluded by {\tt SModelS}. A detailed comparison of the two approaches is on the way. 
We recall, however, that {\tt SModelS} includes in total more than 50 analyses from both ATLAS and CMS, 
which allows to exclude an additional 54\% of points in  {\bf Scan-II}.


\section*{Acknowledgements} 

We thank  D.~Liko for help in using the Vienna Tier-2 Grid, 
and the reactor physics group of LPSC Grenoble for the use of part of their computer resources. 
This work was supported in part by the PEPS-PTI project ``LHC-itools'', 
by the ANR project {\sc DMAstroLHC} and by FAPESP.
U.L.\ and D.P.S.\ are grateful for financial support by the FEMtech initiative of the BMVIT of Austria.
A.L.\ gratefully acknowledges the hospitality of LPSC Grenoble.


\providecommand{\href}[2]{#2}\begingroup\raggedright\endgroup

\end{document}